\documentclass[fleqn, usenatbib]{cls/mnras}

\usepackage[T1]{fontenc}

\DeclareRobustCommand{\VAN}[3]{#2}
\let\VANthebibliography\thebibliography
\def\thebibliography{\DeclareRobustCommand{\VAN}[3]{##3}\VANthebibliography}

\usepackage{booktabs}
\usepackage{pgfplotstable}
\pgfplotstableset{
precision=4,
}
\usepackage{siunitx}
\usepackage{graphicx}
\usepackage{amsmath}
\usepackage{amssymb}
\usepackage{subcaption}
\usepackage{hyperref}
\usepackage{lipsum}
\usepackage{pdflscape}

\newcommand{\draft}{}
\newcommand{\msol}{$\mathrm{M_{\sun}}$}

\newcommand{\cm}{$\mathrm{cm}$}
\newcommand{\barye}{$\mathrm{barye}$}
\newcommand{\K}{$\mathrm{K}$}
\newcommand{\s}{$\mathrm{s}$}
\newcommand{\erg}{$\mathrm{erg}$}
\newcommand{\gcmcubed}{$\mathrm{g}\,\mathrm{cm}^{-3}$}
\newcommand{\kmpersec}{$\mathrm{km}\,\mathrm{s}^{-1}$}

\title[Merger simulations using \textsc{AREPO}]{\textsc{AREPO} White Dwarf merger simulations resulting in edge-lit detonation and run-away hypervelocity companion} 

\author[U. P. Burmester et al.]{
Uri Pierre Burmester$^{1}$\thanks{E-mail: uri.burmester@anu.edu.au (UPB)},
Lilia Ferrario$^{1}$,
R\"udiger Pakmor$^{2}$,
Ivo R.~Seitenzahl$^{3}$,
Ashley J.~Ruiter$^{3}$, and
\newauthor
Matthew Hole$^{1}$
\\
$^{1}$Mathematical Sciences Institute, Australian National University, Canberra ACT 0200, AU\\
$^{2}$Max Planck Institute for Astrophysics, Karl-Schwarzschild-Strasse 1, 85748 Garching, DE\\
$^{3}$School of Science, University of New South Wales, Canberra ACT 2600, AU\\
}

\date{Accepted XXX. Received YYY; in original form ZZZ}

\pubyear{2023}

\usepackage{newtxtext,newtxmath}

\begin{document}
\label{firstpage}
\pagerange{\pageref{firstpage}--\pageref{lastpage}}
\maketitle

\begin{abstract}

We present a series of high-resolution simulations generated with the moving-mesh code \textsc{AREPO} to model the merger of a 1.1~\msol\ carbon-oxygen primary white dwarf with an outer helium layer and a 0.35~\msol\ secondary helium white dwarf. Our simulations lead to detonations that are consistent with the edge-lit scenario, where a helium detonation is ignited at the base of the helium layer of the primary WD, which triggers an off-centre carbon detonation. This produces an asymmetric ejecta pattern and differences in line-of-sight observables (e.g. mean atomic weight). The ejecta that are flung into space are dominated by $^{56}\mathrm{Ni}$, $^{4}\mathrm{He}$, $^{28}\mathrm{Si}$, and $^{32}\mathrm{S}$. Our simulations result in a surviving degenerate companion of mass $0.22-0.25$\,\msol\ moving at $>1\,700$\,\kmpersec, consistent with the observational findings of hypervelocity WDs. The secondary's surface layers are enriched by heavy metals, with  $^{56}\mathrm{Ni}$ making up approximately 0.8\,\% of the remaining mass. We also analyse the sensitivity of the outcome on simulation parameters, including the "inspiral time", which defines a period of accelerated angular momentum loss. We find that the choice of "inspiral time" qualitatively influences the simulation result, including the survival of the secondary. We argue that the shorter inspiral cases result in qualitatively and quantitatively similar outcomes. We also investigate the sensitivity of our results on the primary's chemical profile by comparing simulations using isothermal, constant composition models with the same mass and central composition and characterised by either a bare carbon-oxygen core (no helium) or a carbon-oxygen core enveloped by a thick helium layer.

\end{abstract}




\section{Introduction}
Detailed knowledge of the mechanisms leading to white dwarf (WD) mergers and the outcomes of such mergers have critical downstream effects for our knowledge of many astrophysical phenomena, such as type Ia supernovae \citep[SNe\,Ia,][]{ruiterTypeIaSupernova2020}. SNe\,Ia have a characteristic relationship between the width of their light curve and peak absolute magnitude which allows them to be used as standard candles to measure cosmological distances and the accelerating expansion of the universe \citep{maozObservationalCluesProgenitors2014}. They are also responsible for the production of many rare, heavy elements found in the universe which has enormous influence on models of galactic chemical evolution \citep[GCE, ][]{eitnerObservationalConstraintsOrigin2020}.  Although the progenitor systems and explosion mechanisms that generate SNe\,Ia are still largely unknown \citep[e.g.][]{polinObservationalPredictionsSubChandrasekhar2019, ruiterTypeIaSupernova2020}, there is broad agreement that thermonuclear explosions of carbon-oxygen WDs in close binaries must be involved  \citep{maozObservationalCluesProgenitors2014}. The nature, or the incidence, of the companion star (either another WD or a non-degenerate companion) is not as yet agreed upon by all researchers in this field. Therefore, generating accurate merger simulations to investigate the nature of the progenitor systems of SNE\,Ia and the onset of explosions has a profound impact on our understanding of these astrophysical transients and our models of the chemical composition of our universe \citep[e.g.,][]{eitnerObservationalConstraintsOrigin2022, lachNucleosynthesisImprintsDifferent2020a, kobayashiLowMetallicityInhibitionType1998, eitnerObservationalConstraintsOrigin2020}. This is particularly motivated by the recent discovery of hypervelocity white dwarfs in the Gaia data release DR2 \citep{shenThreeHypervelocityWhite2018}. Simulations are important to constrain the nature, expected range of velocities, and spectra of these objects. 

However, simulating such mergers is not a trivial affair. During a merger, gasses from the two stars interact in a 3D space with the temperature, pressure and position set by the complex interaction of fluid flow, and chemical composition. From a numerical standpoint, the values of variables like density vary across a dozen orders of magnitude and interactions can take place over both large and small length scales, which necessitates careful consideration of numerical accuracy. A key question is when and under which conditions a detonation occurs as result of binary interaction, and what effect our choice of numerical scheme has on the outcome \citep{seitenzahlSpontaneousInitiationDetonations2009, seitenzahlInitiationDetonationGravitationally2009}.

We describe the codes we use for our merger simulations in Section~\ref{sec:method} and the specific parameters of the simulations in Section~\ref{sec:setup}. We show the results of the simulations and the stability of the new structures in Section~\ref{sec:results}, and the effects of spiral-in acceleration in Section~\ref{sec:inspiral}. Finally we summarise our results in Section~\ref{sec:conclusions}.

\section{Prior Work} \label{sec:prior}
Thermonuclear detonations of carbon–oxygen (CO) WDs have been suspected progenitors of SNe Ia for decades \citep{hoyleNucleosynthesisSupernovae1960}. Several possible scenarios leading to a SN\,Ia explosion are envisaged, with significant overlap between the various categories. These scenarios can be listed as follows: (a) The double degenerate (DD) scenario, based on the merger of two WDs. This scenario may lead to either collapse or explosion as first reported by \citet{ibenSupernovaeTypeEnd1984, webbinkDoubleWhiteDwarfs1984}. More recently, \citet{pakmorHeliumignitedViolentMergers2013} discusses violent mergers as a possible sub-channel of the DD scenario, while others consider very long delays from merger to explosion, due to the rapid rotation of the resulting object that keeps it ``overstable'' above the classical Chandrasekhar limit \citep{tornambePreexplosiveObservationalProperties2013}. (b) The single degenerate (SD) scenario \citep[e.g.][]{hanSingledegenerateChannelProgenitors2004, whelanBinariesSupernovaeType1973}. Here a WD accretes mass from a non-degenerate stellar companion until its interior reaches a critical density beyond which explosive $^{12} \mathrm{C}$ burning occurs \citep[e.g][]{gasquesNuclearFusionDense2005,seitenzahlThreedimensionalDelayeddetonationModels2013}. (c) The core-degenerate (CD) scenario \citep[e.g.][]{sokerCircumstellarMatterSupernova2015, sparksSupernovaResultDeath1974}. Here the WD merges with the hot core of a massive asymptotic giant branch (AGB) star. In this case the explosion might occur immediately or a long time after the merger has occurred because of the rapid rotation of the resulting object. (d) The ``double-detonation'' mechanism \citep{livneGeometricalEffectsOffCenter1990}. Here, a sub-Chandrasekhar mass WD accumulates a layer of helium-rich material accreted from a helium-rich donor star. The helium layer on the primary is compressed as more material is accreted until it detonates, leading to a second (carbon) detonation in the core \citep{gronowDoubleDetonationsSubMCh2021, finkDoubledetonationSupernovaeSubChandrasekhar2007}. (e) The WD–WD collision scenario \citep[e.g.][]{kushnirHeadonCollisionsWhite2013, hutGlobularClusterEvolution1985}. In this scenario either a third star brings two WDs to collide, or the dynamical interaction occurs in a dense stellar system (e.g., a globular cluster), where such interactions are more likely to occur. In some cases, the collision results in an immediate explosion. Despite some attractive features of this scenario, it could account for at most a few per cent of all SNe Ia \citep{hallakounLimitsPopulationCollisionaltriples2019}.

Currently, double-degenerate systems are the favourite candidates to be progenitors for multiple reasons: (i) their numbers are more consistent with the observed rate of supernovae \citep{ruiterRatesDelayTimes2009}, (ii) because they are very dim, they cannot be easily detected prior to the explosion in electromagnetic light, and (iii) lack of any surviving companion brighter than about solar luminosity \citep{pakmorFateSecondaryWhite2022}. \citet{shenNonlocalThermodynamicEquilibrium2021} has also shown that simulations of non-LTE radiative transfer of sub-Chandrasekharmass WD detonations are able to reproduce the entirety of the Phillips relation \citep{phillipsAbsoluteMagnitudesType1993}, from subluminous to overluminous SNe Ia.

Over the last few years, support has increased that some explosions ignite via the double detonation mechanism outlined by \citet{livneSuccessiveDetonationsAccreting1990}, with particular interest the ``helium-ignited violent merger'' or ``dynamically-driven double-degenerate double-detonation'' (D6) models . Corresponding to (d) above, the mechanism is understood as follows: unstable dynamical accretion of helium from the secondary, less massive, WD just prior to the merger of the binary heats up the helium shell on the primary, more massive, WD. Eventually, as the accreted layer grows in mass, the base of the helium layer becomes sufficiently hot and dense that a helium detonation may be ignited. Under some circumstances, the He-detonation can directly transition into a carbon detonation as the leading shock-wave of the detonation penetrates into the core. This scenario is called the ``edge-lit'' detonation case \citep{forcadaEdgelitDoubleDetonation2007,livneGeometricalEffectsOffCenter1990, yungelsonSupernovaRatesCosmic2000}.  Alternatively, the helium detonation could cause burning across the full helium shell around the primary sending a shock-wave into its core. The shock-wave then converges into a single point in the carbon-oxygen core causing its detonation \citep{livneSuccessiveDetonationsAccreting1990,pakmorHeliumignitedViolentMergers2013, shenIgnitionCarbonDetonations2014}. In the double detonation scenario the less massive WD is generally believed to survive, because it is still intact when the more massive WD explodes (excluding possibly the exploding secondary from \citealt{pakmorFateSecondaryWhite2022}). The existence of surviving secondaries is supported by the discovery of hyper-velocity WDs by \citet{shenThreeHypervelocityWhite2018} in the DR2 Gaia data release. Some work has also been done to characterise the expected ejecta morphology of a double detonation. \citet{ferrandDoubleDetonationDouble2022} investigate a D6 supernova model and carry it into the supernova remnant (SNR) phase up to 4000 years after the explosion, past the time when all the ejecta have been shocked. The first detonation produces an ejecta tail visible at early times, while the second detonation leaves a central density peak in the ejecta that is visible at late times. The SNR shell is off-centre at all times, because of an initial velocity shift due to binary motion. The companion WD produces a large conical shadow in the ejecta, visible in projection as a dark patch surrounded by a bright ring \citep{ferrandDoubleDetonationDouble2022}.

A constant composition, isothermal WD structure is usually generated by choosing a central density and temperature based on the desired total mass of the WD.  Then a numerical integrator is employed that uses the equations of hydrostatic equilibrium to generate the pressure and interior mass for a set of radii. The value of the other primitive variables, such as density, are then set by either assuming a power-law relationship (i.e. a polytrope) or by using an equation of state (EOS).

\begin{table}
\centering
\caption{The 55-species nuclear reaction network list which is coupled to \textsc{AREPO}, used in \citet{pakmorThermonuclearExplosionMassive2021}.}
\label{tab:network}
\begin{tabular}{|l|l|l|l|l|}
El.  &  .  &  .  &  . &  .  \\ \hline
n   & O17  & Mg26 & S32  & Sc43 \\
p   & F18  & Al25 & S33  & Ti44 \\
He4 & Ne19 & Al26 & Cl33 & V47  \\
B11 & Ne20 & Al27 & Cl34 & Cr48 \\
C12 & Ne21 & Si28 & Cl35 & Mn51 \\
C13 & Ne22 & Si29 & Ar36 & Fe52 \\
N13 & Na22 & Si30 & Ar37 & Fe56 \\
N14 & Na23 & P29  & Ar38 & Co55 \\
N15 & Mg23 & P30  & Ar39 & Ni56 \\
O15 & Mg24 & P31  & K39  & Ni58 \\
O16 & Mg25 & S31  & Ca40 & Ni59
\end{tabular}
\end{table}

There are many combinations of WDs (e.g. He, `hybrid' HeCO, CO, ONe) that can be found in a binary system. There is also a continuum of possible other factors that influence the nature of these mergers, such as mass ratio angular momentum. What remains unclear is which progenitors correspond to which observable outcomes (e.g. SNe\,Ia, R Coronae Borealis stars, and neutron stars from merger induced collapse). This is the domain of parameter scan studies \citep[e.g., see for example][]{shenEveryInteractingDouble2015}.

Several papers have made contributions in this field recently. \citet{danHowMergerTwo2012} considered the effect of the mass ratio on the orbital stability and detonation characteristics of WD-WD systems. The group investigated a parameter space of 200 simulations with masses ranging from 0.2 to 1.2\,\msol, using a 3D Smoothed Particle Hydrodynamics (SPH) code, Helmholtz EOS, and various chemical profiles. \citet{danHowMergerTwo2012} find excellent agreement with the orbital evolution predicted by mass transfer stability analysis and find that a large fraction of He-accreting WDs should explode. They find that all dynamically unstable systems with accretor masses below 1.1\,\msol \, and donor masses above $\approx 0.4$\,\msol \, trigger a helium detonation at surface contact. \citet{roy3DHydrodynamicalSimulations2022} use 3D hydrodynamical simulations to explore this case in which a helium detonation occurs near the point of Roche lobe overflow of the donor WD and may lead to an SN Ia through the D6 mechanism. They find that the helium layer of the accreting primary WD does undergo a detonation, while the underlying carbon-oxygen core does not, leading to an extremely rapid and faint nova-like transient instead of a luminous SN Ia event. \citet{tanikawaDoubledetonationModelsType2019} also conduct high-resolution smoothed particle hydrodynamic (SPH) simulations. The supernova ejecta of the primary white dwarfs strip materials from the companion white dwarfs, whose mass is approximately $10^{-3}$ solar masses. The stripped materials contain carbon and oxygen when the companion white dwarfs are carbon-oxygen white dwarfs with helium shells less than or equal to 0.04 solar masses. 

An ideal merger simulation would follow the progenitor objects from a time long before any gravitational interaction and would accurately simulate the gravitational forces and nuclear reactions in real-time. In reality, it is not computationally feasible to simulate the merger beginning from Roche-lobe overflow until the actual merger occurs -- the timescale for the removal of angular momentum via gravitational radiation losses is much longer than the dynamical timescale of the simulation. A solution which does not sacrifice much physical complexity is to remove angular momentum at an artificially-high rate. \citet{pakmorThermonuclearExplosionMassive2021} describes an \textsc{INSPIRAL} method that consists of adding a tidal force that decreases the separation of the stars at a constant rate. The importance of this inspiral aspect and the sensitivity of the simulation results on the specifics will be further discussed later. 

Ideally, a study of merging WDs will include high-resolution simulations of different WD masses and chemical composition profiles. We defer a more complete parameter scan to future work. For this paper, we explore the effect of including more realistic chemical profiles for the WDs (derived from stellar evolution calculations) compared to ad-hoc compositions. We present our studies of the merger of a binary system with a 1.1\,\msol \, CO primary and a 0.35\,\msol \, He secondary. We have chosen this combination of masses because it lies on the boundary of different supernovae models in previous studies \citep[e.g., see Figure 3 in][]{shenEveryInteractingDouble2015}. 

In this paper we seek to answer the following questions (1) Does a detonation occur? (2) Do the simulation outcomes vary sensitively on the chemical profile of the progenitors? (3) Do the simulation outcomes vary sensitively on the inputs to the ``inspiral'' phase? (4) What is the chemical yield of the detonation and does the secondary WD survive?

\section{Methods} \label{sec:method}

We explore the dependence of the simulation outcomes on changes in the composition profile. Specifically, we compare models of the commonly used isothermal, homogeneous composition profile to a model obtained using the White Dwarf Evolution Code (WDEC) \footnote{\url{https://github.com/kim554/wdec}}. While the latter is a single star evolution-based profile, it nevertheless allows us to investigate the role of the chemical composition and structure profile on the outcome of a merger.

\subsection{White Dwarf Evolution Code}
One of the primary motivations to use a WD modelling code is to explore the effects of non-homogeneous chemical abundance distributions. While constant composition models are largely accurate to first order (e.g. total mass, density, and temperature of the interior), they neglect some crucial physics concerning the internal chemical structure of WDs. However, because computations of temperature and nuclear reaction rates are highly dependent on chemical composition, we believe it is more appropriate to include a more realistic WD structure. We have chosen to use WDEC, which has a long lineage in WD research and makes use of the latest physics to give a more accurate structure. WDEC is written in \textsc{FORTRAN} and has the ability to generate structures across a wide range of masses. The code evolves hot ($\approx 100,000$\,\K) input models down to a chosen effective temperature by relaxing the models as solutions of the equations of stellar structure. WDEC was reworked by \citep{bischoff-kimWDECCodeModeling2018} to allow for the use of Modules for Experiments in Stellar Astrophysics (MESA) equations of state and opacity tables (version 8118) \citep{paxtonModulesExperimentsStellar2019}. 

\subsection{\textsc{AREPO}}

The \textsc{AREPO} code \citep{springelPurSiMuove2010,pakmorImprovingConvergenceProperties2016a,weinbergerAREPOPublicCode2020} uses a second-order finite-volume method to solve the ideal MHD equations on an unstructured grid. A Voronoi mesh is generated in each timestep from a set of mesh-generating points that move along with the flow, thus ensuring a nearly Lagrangian behaviour while regularising the mesh. The fluxes over the cell boundaries are computed using a Harten-Lax-van Leer discontinuities (HLLD) approximate Riemann solver \citep{pakmorMagnetohydrodynamicsUnstructuredMoving2011a}. \textsc{AREPO} solves self-gravity with a one-sided octree solver. \textsc{AREPO} was used to evaluate the dynamics of the binary because it is able to handle the full 3D evolution of a large number of points and has built-in support for evaluating equations of state for degenerate matter. While \textsc{AREPO} is capable of simulating the effects of magnetic fields, we do not use these capabilities in these purely hydrodynamic simulations.

\begin{figure*}
\centering
\includegraphics[width=\textwidth, \draft]{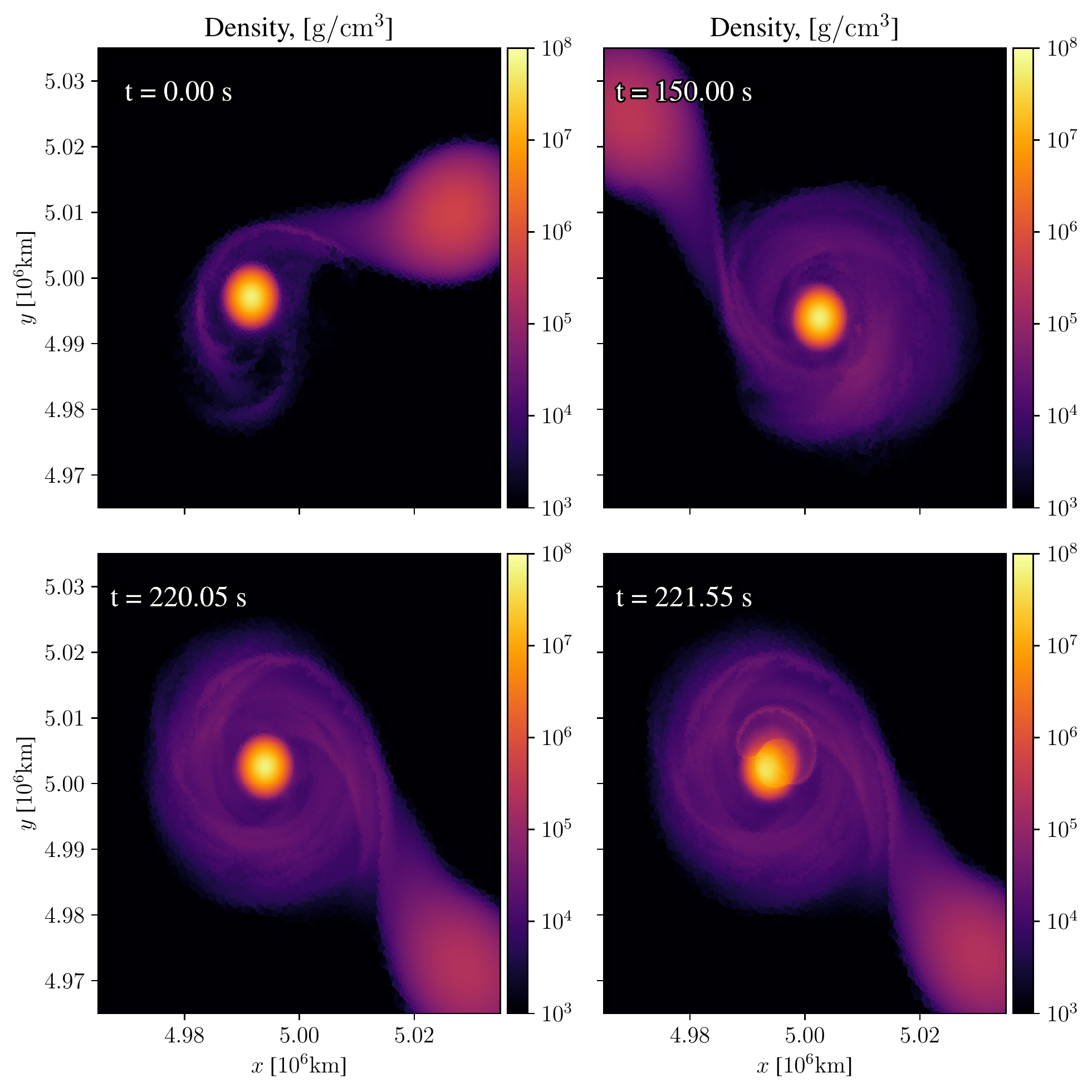}
\caption{Density time evolution of the merging event for T-120. (a) At the beginning of the merger phase, the accretion stream begins to form as the secondary overflows its Roche Lobe. (b) An accretion disk begins to form. (c) The accretion disk grows in thickness. (d) The He detonation spreads around the sides of the primary while the degenerate interior is converted to mostly $^{56}\mathrm{Ni}$.}
\label{fig:time_evolution}
\end{figure*}

\begin{table*}
\centering
\caption{A summary of the inspiral cases explored, where the model name indicates the inspiral time in seconds (e.g. T-120 includes 120 seconds of inspiral time). $\mathrm{a_2}$ and $\mathrm{T_2}$ are the orbital separation and period at the end of Phase 2 respectively. $\mathrm{t_3}$ specifies the amount of time in Phase 3, i.e. from the end of the inspiral to detonation, if any. $\mathrm{M_{sec, surv}}$ and $\mathrm{v_{sec, surv}}$ refer to the mass and velocity of the surviving secondary, determined by the amount of bound mass at a time 100 seconds after detonation. $\mathrm{\Delta_{det}}$ is the time difference between the onset of Helium and Carbon detonation.}
\label{tab:inspiral-comparison}
\resizebox{\linewidth}{!}{%
\begin{tabular}{|p{0.1\linewidth}|p{0.1\linewidth}|p{0.15\linewidth}|p{0.2\linewidth}|p{0.25\linewidth}|p{0.15\linewidth}|p{0.1\linewidth}|}
\hline
Model & $\mathrm{a_2 [km]}$ & $\mathrm{T_2}$ & $\mathrm{t_3 [s]}$ & $\mathrm{M_{sec, surv}}$ [\msol] & $\mathrm{v_{sec, surv}}$ [\kmpersec] & $\mathrm{\Delta_{det}}$ [\s] \\ \hline
T-150 & 32,900 & 85.6 & 115 & 0.00 & N/A  & 0.2 \\ \hline
T-135 & 34,600 & 92.1 & 130 & 0.15 & 1790 & 0.2 \\ \hline
T-120 & 35,500 & 95.7 & 221 & 0.22 & 1840 & 0.3 \\ \hline
T-115 & 35,600 & 96.4 & 231 & 0.25 & 1700 & 0.3 \\ \hline
T-113 & 35,700 & 96.5 & 222 & 0.23 & 1710 & 0.3 \\ \hline
T-110 & 35,700 & 96.6 & N/A & N/A  & N/A  & N/A \\ \hline
\end{tabular}%
}
\end{table*}

\begin{figure*}
    \centering
    \includegraphics[width=0.7\linewidth, \draft]{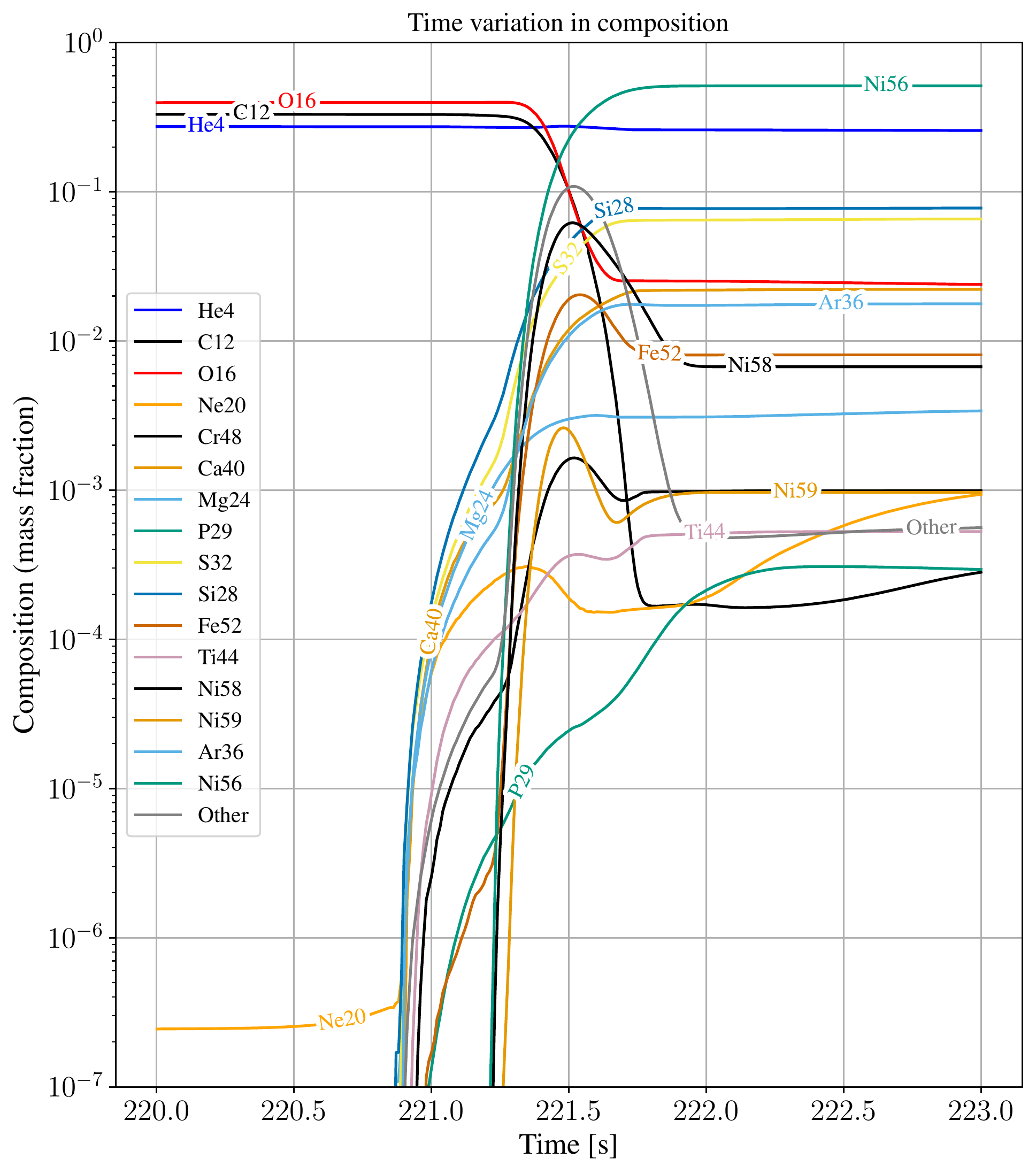}
    \caption{Time-variation of the composition of all simulated mass (primary \& secondary) for case T-120. The $y$-axis gives the percentage of mass made up by a specific species (e.g. at late times, $^{56}\mathrm{Ni}$ makes up 51.25\,\% of the mass). ``Other'' indicates the summation of all species in the network not listed in the legend. Note that the proportion of higher elements increases dramatically at $t = 220.9$\,s and $t = 221.2$\,s, indicated by groups of near-vertical lines at these times.}
    \label{fig:composition}
\end{figure*}

\subsection{Setup \label{sec:setup}}
In this experiment, we simulate a DD merger involving a primary CO WD with a He layer (we shall call this type of object a COHe WD), a mass of 1.1\,\msol\ and an effective temperature of $40,000$\,\K. The mass of the primary WD's He layer was set to be approximately 2 per cent of the total mass, consistent with observations of isolated WDs  \citep{mccookCatalogSpectroscopicallyIdentified1999}. 

Having created a 1D stellar structure for the massive primary using WDEC, we then ported this structure into \textsc{AREPO}. A HEALPIX (Hierarchical Equal Area isoLatitude Pixelisation) mapping \citep{gorskiHEALPixFrameworkHighResolution2005} was used to transform this 1D structure to 3D. This mapping treats every point in the one-dimensional profile as a radius and generates a series of points equally spaced in latitude and longitude. This creates a spherically symmetric structure that matches the output of WDEC. 

We employ explicit refinement and de-refinement when the mass of a cell is larger than twice or smaller than half of the target mass resolution (set to $10^{-6}$\,\msol \, for all simulations in this paper). Additional refinement is triggered when the volume of a cell is more than 10 times larger than its smallest direct neighbour to avoid large resolution gradients in the mesh at steep density gradients. Moreover we enforce a maximum volume for cells of $10^{30}$\,cm$^3$ to prevent de-refinement of the background mesh. We soften the gravitational force to avoid spurious two-body interactions with a softening length of 2.8 times the radius of a cell, but force the softening to be at least 10\,km.

The values of primitive variables are computed using a non-ideal Helmholtz equation of state (HES) \citep{timmesAccuracyConsistencySpeed2000}.  Moreover, we fully couple a 55-isotope nuclear reaction network \citep{pakmorThermonuclearExplosionMassive2021} to the simulation using the Joint Institute for Nuclear Astrophysics \textsc{REACLIB} database\footnote{\url{http://groups.nscl.msu.edu/jina/reaclib/db/}}. A list of species present in the network is shown in Table~\ref{tab:network}. The nuclear reaction network is active for all cells with $T > 10^6$\,\K, but excludes shocked cells which we define as cells with $\nabla \cdot \vec{v} < 0$ and $|\nabla P| r_{cell} / P < 0.66 $ for numerical stability, where $\vec{v}$ is the cell's velocity, $r_{cell}$ is the cell's radius, and $P$ is the pressure \citep{seitenzahlSpontaneousInitiationDetonations2009}. Hydrogen was deliberately excluded from the WDEC atmospheric model because it would be a very small portion of the total mass and difficult to resolve at the current mass resolution,  has little bearing on the final chemical profile, and would make the simulations unnecessarily too time-consuming. 

To create a structure for the low-mass secondary WD, we use the isothermal, constant composition approximation. Thus, we choose a temperature of $T = 5 \times 10^5$\,\K \, and a pure helium composition. Its structure is generated using a simple numerical integrator as outlined in \citet{pakmorFateSecondaryWhite2022}. This involves estimating the central density of the star using the total mass and then integrating the equations of hydrostatic equilibrium to a cutoff density of $\rho_c = 1 \times 10^{-4}$\,\gcmcubed.

Our simulation consists of a three-step process: (1) Relaxation, (2) Inspiral, (3) Merger Stage.  Step (1) is needed because the WD structure may have suffered from discretisation errors in the transformation from 1D to 3D which disrupt hydrostatic equilibrium. Thus, a relaxation method is employed to create stable stellar models. We damp away spurious velocities for both WDs in isolation for 10\,s (i.e. several dynamical timescales) using the criterion outlined in \cite{ohlmannConstructingStable3D2017} and then evolve for approximately 100\,\s \, to confirm stability of the stellar structure, whilst checking for energy conservation. 

In step (2), we set the two WDs on a circular orbit with an initial orbital period of $T = 80$\,\s \, and a separation of $a = 4.25 \times 10^9$\,cm. This separation is large enough that the WDs are essentially undisturbed when they feel the gravitational force exerted by their companion. The simulation box has a side length of $10^{12}$\,cm. We fill the background mesh with a density of $10^{-4}$\,\gcmcubed \, to avoid numerical problems arising from steep gradients from the stellar surface to empty space while still only adding a negligible amount of mass. At this phase we also add so-called ``Passive Scalar Labels'' to track the progress of the simulation. This involves labelling all cells which initially belonged to the primary and the secondary. These labels allow us to perform diagnostics such as averaging the position of the interior particles to estimate the location of the WD's centre.

Our inspiral technique is the same as that adopted in \citet{pakmorThermonuclearExplosionMassive2021}, with enhanced gravitational radiation active for approximately 1.5 orbits, decreasing the separation at a constant rate of $50$\,\kmpersec. When referring to this algorithm, as opposed to the simulation phase, we use the term \textsc{INSPIRAL}. Several different simulations were run using the same two progenitor objects and different lengths of time during the inspiral phase. A summary of these cases is shown in Table~\ref{tab:inspiral-comparison}. The inspiral times $T_I$ were chosen by experimentation focused on their effects on the properties of the binary. $T_I \leq 110\,\mathrm{s}$ resulted in simulations with low mass transfer rates which would not conclude in over 500,000 timesteps, beyond which time aggregated errors become an issue. $T_I \geq 150$\,\s, however, resulted in a complete disruption of the secondary before even one orbit had been completed. Experimentation within the remaining (reasonable) range revealed that the duration of this accelerated inspiral phase influences the survival of the secondary, its mass, and associated observables, but not the physical outcome of the merger (see Section~\ref{sec:inspiral}). The case we will mainly discuss in this paper is T-120, when we stop the accelerated inspiralling at $T = 120$\,\s. At this point the binary's orbit has shrunk to a separation $a = 3.5 \times 10^9\,\mathrm{cm}$. From here onward the total angular momentum is  conserved for the rest of the simulation. Step (3) involves running \textsc{AREPO} in its normal mode until either detonation or accretion-induced collapse occur and with the secondary either destroyed or flung away.

\section{Simulation Results} \label{sec:results}

\subsection{Isolated Phase and WDEC structure stability} \label{sec:stability}

To be confident in the stability of the structure given by WDEC in 3D (under conditions like, e.g.,  Hydrostatic Equilibrium), we examine the differences in structure before and after mapping. This was done using radial profiles from WDEC and samples from the \textsc{AREPO} output at various times throughout the merger evolution. We determined that the structure is stable, with some expected ``smearing'' of the low-mass regions at large radii which is expected given the orders of magnitude difference between the stellar interiors and the low-density background ($\rho \leq 10^3$\,\gcmcubed). At large radii there is only a handful of cells available to model the low-density region and they are very large. 

\begin{figure*}
    \centering
    \includegraphics[width=\linewidth, \draft]{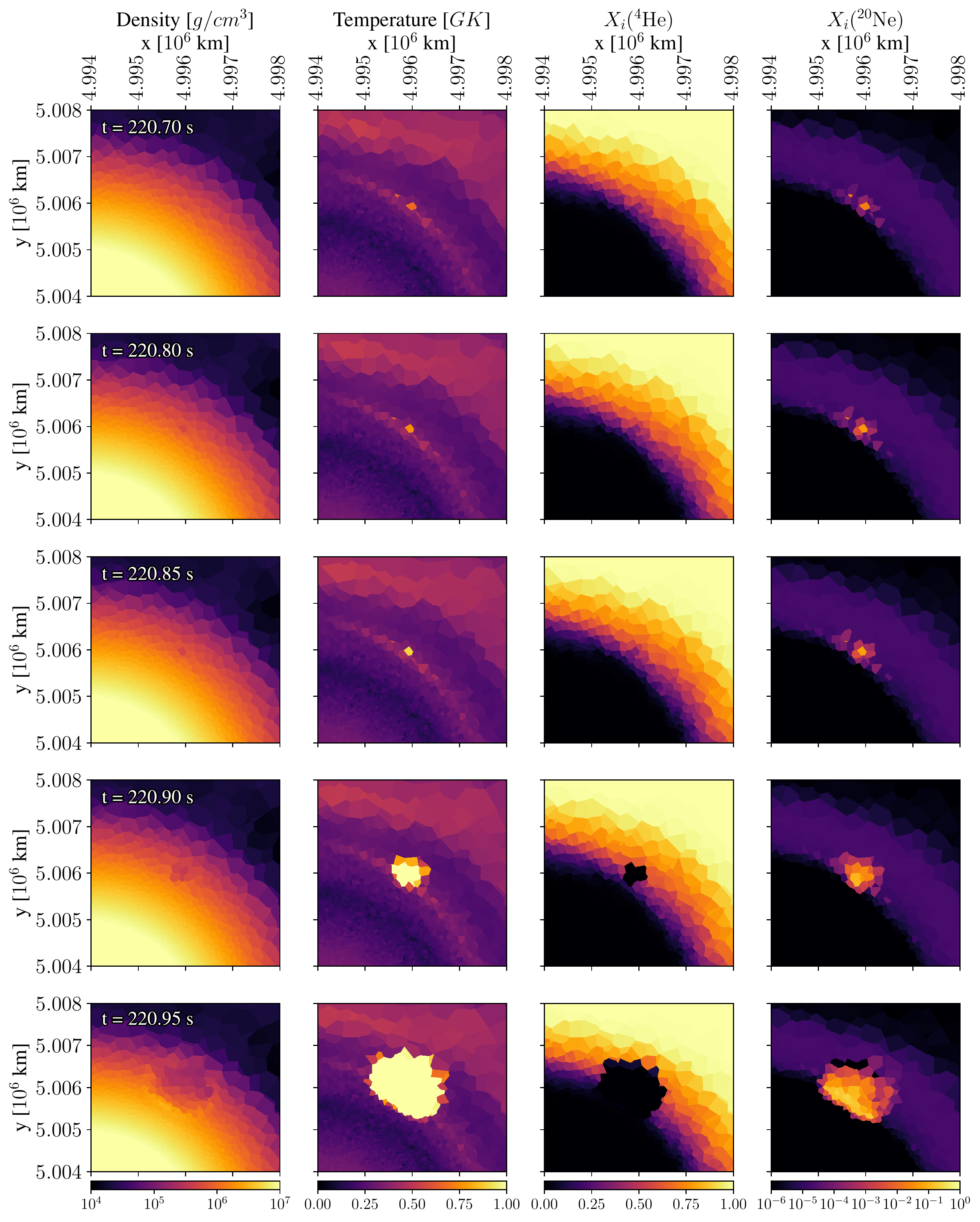}
    \caption{Evolution snapshots of primitive variables around the time of He ignition. Note how the disturbance begins at the bottom of the helium layer, in the mixed C-O-He region. This is preceded by the production of Neon in the same area.}
    \label{fig:detonation_pcolor}
\end{figure*}

\begin{figure*}
    \centering
    \includegraphics[width=\linewidth, \draft]{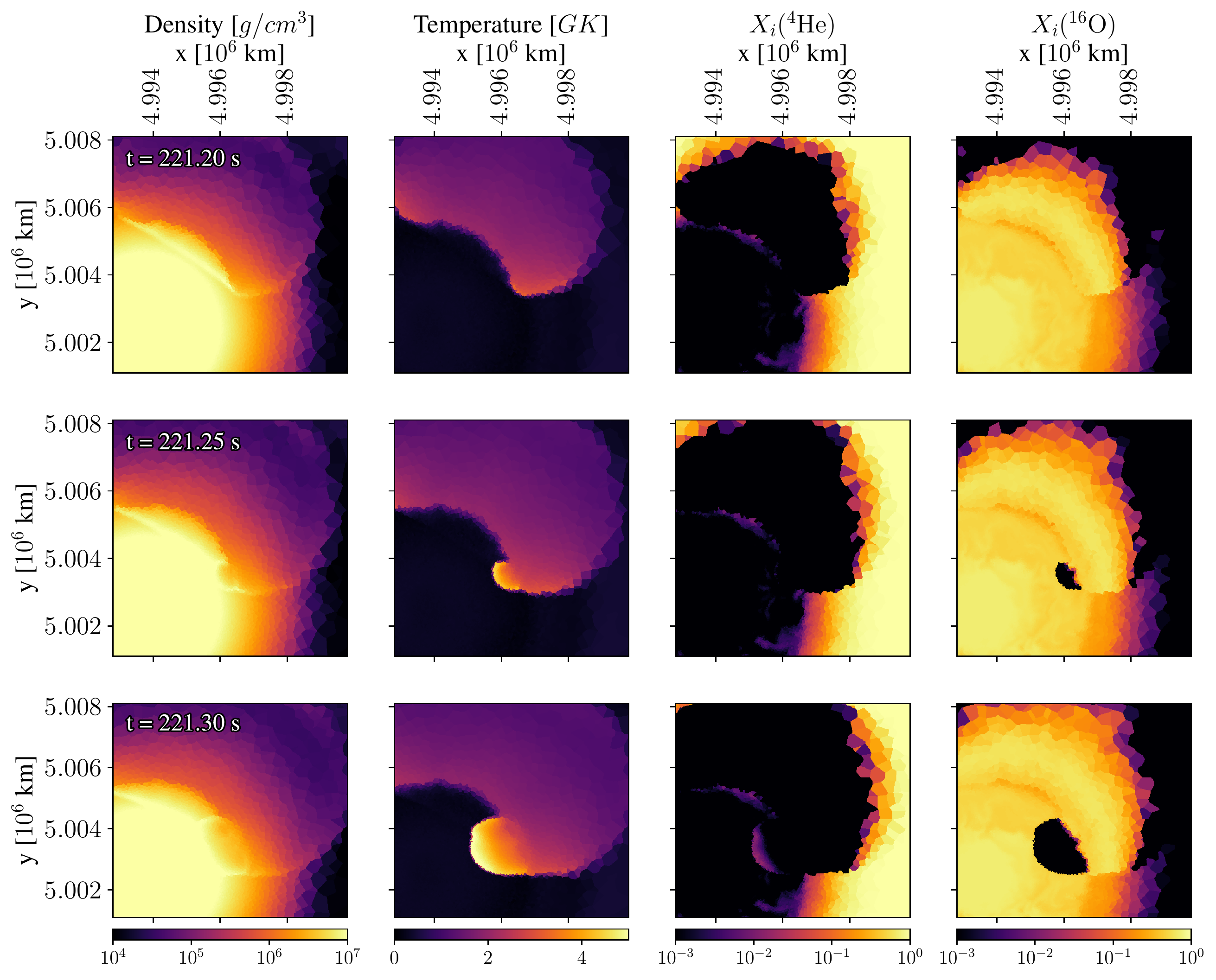}
    \caption{Location of elements at initiation of carbon detonation. The depletion of oxygen at $t = 221.5\,\mathrm{s}$ is represented by the growing black region in the centre.}
    \label{fig:xnuc_pcolor}
\end{figure*}

\subsection{Inspiral, Evolution and Merger} \label{sec:detonation}

Figure~\ref{fig:time_evolution} shows the time evolution of the merger using a series of snapshots of the density of the cells. In the upper left frame ($t=0$), which is taken at the end of the inspiral phase, we can see that a small accretion stream has begun to form and limited mass exchange is taking place. Due to the relatively high initial angular momentum of the system, accretion occurs slowly over a period of several hundred seconds. During this time, the two objects continue to orbit around each other and mass continues to accrete onto the primary. During this period, nuclear reaction rates are marginal, so we use only a 13-species subset of the 55-species network from Table~\ref{tab:network}. This includes: $^{4}\mathrm{He}$, $^{12}\mathrm{C}$, $^{16}\mathrm{O}$, $^{20}\mathrm{Ne}$, $^{24}\mathrm{Mg}$, $^{28}\mathrm{Si}$, $^{32}\mathrm{S}$, $^{36}\mathrm{Ar}$, $^{40}\mathrm{Ca}$, $^{44}\mathrm{Ti}$, $^{48}\mathrm{Cr}$, $^{52}\mathrm{Fe}$, and $^{56}\mathrm{Ni}$. The following two frames of Figure~\ref{fig:time_evolution} show the normal evolutionary process; the accretion disk continues to grow in size as the secondary experiences Roche Lobe overflow and the accumulating matter increases the pressure and density on the primary. In the lower-left frame the accretion disk consists of approximately 0.13\,\msol. In the lower-right figure, we can see a horseshoe-shaped slice of a helium-detonation region extending halfway around the primary; there is also a high-density region in the interior which is distinct from the rest of the primary and is caused by a carbon detonation. We must now examine the triggers of these events using a finer temporal resolution, repeating these timesteps with the 55-species network.

Figure~\ref{fig:composition} sheds some light on the timing of the various events; this figure shows the proportion of mass in the entire simulated box made up by each of the species listed in Table~\ref{tab:network}. Prior to $t = 220.5$, the binary contains nearly the same quantities at almost the same proportions of $^{4} \mathrm{He}$, $^{12} \mathrm{C}$ and $^{16}\mathrm{O}$ as at the start of the simulation, which marginal amounts of $^{20}\mathrm{Ne}$, $^{24}\mathrm{Mg}$ and $^{28}\mathrm{Si}$ produced. The Neon production is predicted by \citet{shenIgnitionCarbonDetonations2014} as a result of a $^{16} \mathrm{O} (\alpha, \gamma) ^{20} \mathrm{Ne}$ process. At the temperatures and densities in the mixed Carbon-Oxygen-Helium region, this dominates over other processes such as Triple-$\alpha$ \citep[See Figure 1]{shenIgnitionCarbonDetonations2014}. However, at approximately $t = 220.9$\,s, the production rate of these elements increases dramatically. Approximately $0.3$\,s \, later, a second detonation begins which increases production of $^{58}\mathrm{Ni}$, $^{56}\mathrm{Ni}$ and other elements. After a period of several seconds of fast changes in composition, we can see that the proportions stabilise as the primary is disrupted and flung into space as heavy ejecta. The yield of these two phases of transformation is shown in Table~\ref{tab:yield}, where the ``He det'' and ``C det'' columns indicate the difference in proportion of each of the network species before and after each of the transformation phases. Note that these yields are based on time rather than location. For example, helium detonation is likely to continue while carbon detonation is in progress. Since neither Figure~\ref{fig:composition} nor Table~\ref{tab:yield} contains this position information, we require snapshots which show the relevant physical properties of the locations where He and C are triggered.

Figure~\ref{fig:detonation_pcolor} shows the dynamics where He is triggered at around $t = 220.9$\,s, allowing us to see the influence of the individual species on the detonation dynamics. We can see that the detonation is triggered at the base of the helium layer at $r = 4.0 \times 10^8$\,cm, in the mixed C-O-He region of the primary. This base is bounded by the black interior region deficient of helium. Examining the neighbourhood of cells around this point, we can determine that the approximate value of the primitive variables just prior to the He detonation are: $\rho = 9.0 \times 10^5$\,\gcmcubed, $p = 3.3 \times 10^{22}$\,\barye, $T = 5.9 \times 10^{8}$\,\K. The chemical composition in this neighbourhood is: $ X(^{12}\mathrm{C}) = 0.43, X(^{16}\mathrm{O}) = 0.08, X(^{4}\mathrm{He}) = 0.49$. The first rows of the image show that these radii are an ideal location for the production of $^{20}\mathrm{Ne}$ and this serves as a useful precursor to the detonation (note how the high Neon production is mirrored in a high-temperature region at $t = 220.70$). 

In our simulation the helium detonation does not completely encircle the core of our high-mass primary; instead, carbon detonation is triggered when the helium detonation raises the temperature and density to $T\ge 3.5\times10^9$\,K,  $\rho\approx 6.8\times10^6$\,g\,cm$^{-3}$ in a cell at radius $r = 2.87 \times 10^8$\,cm. This is significantly in excess of the minimum requirements for carbon detonation (i.e. $T\ge 2.5\times10^9$\,K,  $\rho\approx 2\times10^6$\,g\,cm$^{-3}$; \citealt{seitenzahlInitiationDetonationGravitationally2009}). The chemical composition just prior to carbon detonation is: $ X(^{12}\mathrm{C}) = 0.48, X(^{16}\mathrm{O}) = 0.52, X(^{4}\mathrm{He}) < 1 \times 10^{-3}$. Figure~\ref{fig:xnuc_pcolor} shows the location of various elements being produced during the helium detonation stage and the first penetration into the interior. Noting the location of $^{28}\mathrm{Si}$, for example, we see how this second detonation extends inwards from the helium detonation region. The detonation of the helium shell then compresses the CO core, causing an off-centre detonation which completely destroys the primary. 

The two-phase detonation we observe in Figures~\ref{fig:detonation_pcolor} and \ref{fig:xnuc_pcolor} is most consistent with the characteristics of the edge-lit scenario (d) in Section~\ref{sec:prior}: the helium detonation occurs at the base of the helium layer and proceeds around the core for approximately 0.3\,\s. However, the helium detonation is not able to completely encircle the core. The relatively short time difference between the two detonations together with a high-mass, energetic primary are also consistent with edge-lit scenario. This outcome is consistent with the results shown in Figure 1 of \citet{danStructureFateWhite2014}, who predicted a double detonation scenario for a (1.1, 0.35) \msol \, binary pair. 

\subsection{Ejecta} \label{sec:yield}

\begin{figure*}
    \centering
    \includegraphics[width=0.8\linewidth, \draft]{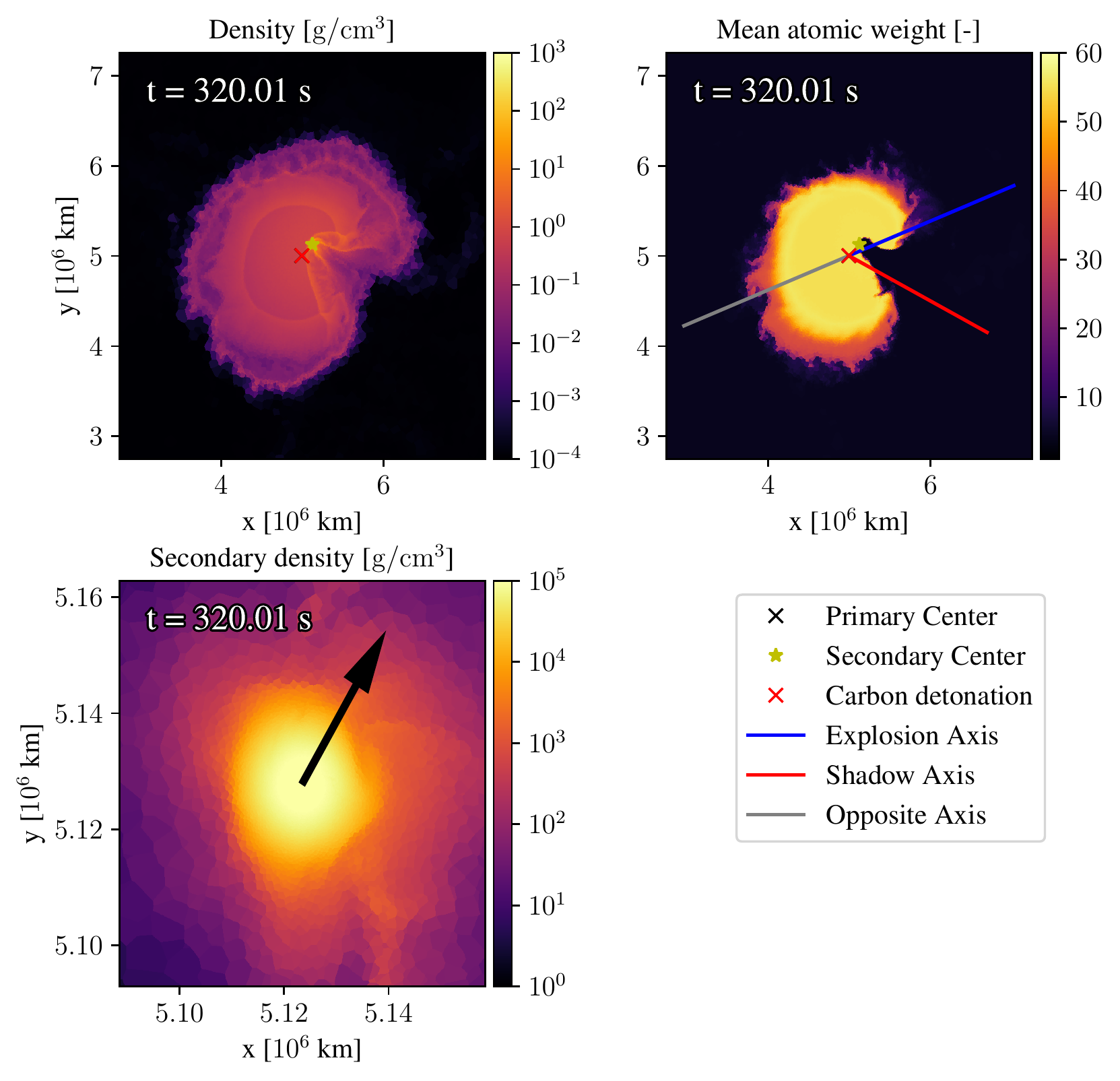}
    \caption{Expansion of ejecta cloud approximately 100\,s after the time of explosion. At this point, the ejecta can be considered to be in homologous expansion; the interior of the star has been uniformly converted to $^{56}\mathrm{Ni}$, causing the solid yellow region on the mean atomic weight plot in Fig.\ref{fig:maw_axis}. Intermediate species can be found in between this Ni bubble and the fast-moving, low mass He front. The coloured rays correspond to the axes in Figure~\ref{fig:maw_axis}. Note that the location of the ``Primary Centre'' (prior to explosion) and ``Carbon detonation'' are indistinguishable on this length-scale.}
    \label{fig:ejecta}
\end{figure*}

\begin{figure*}
    \centering
    \includegraphics[width=0.8\linewidth, \draft]{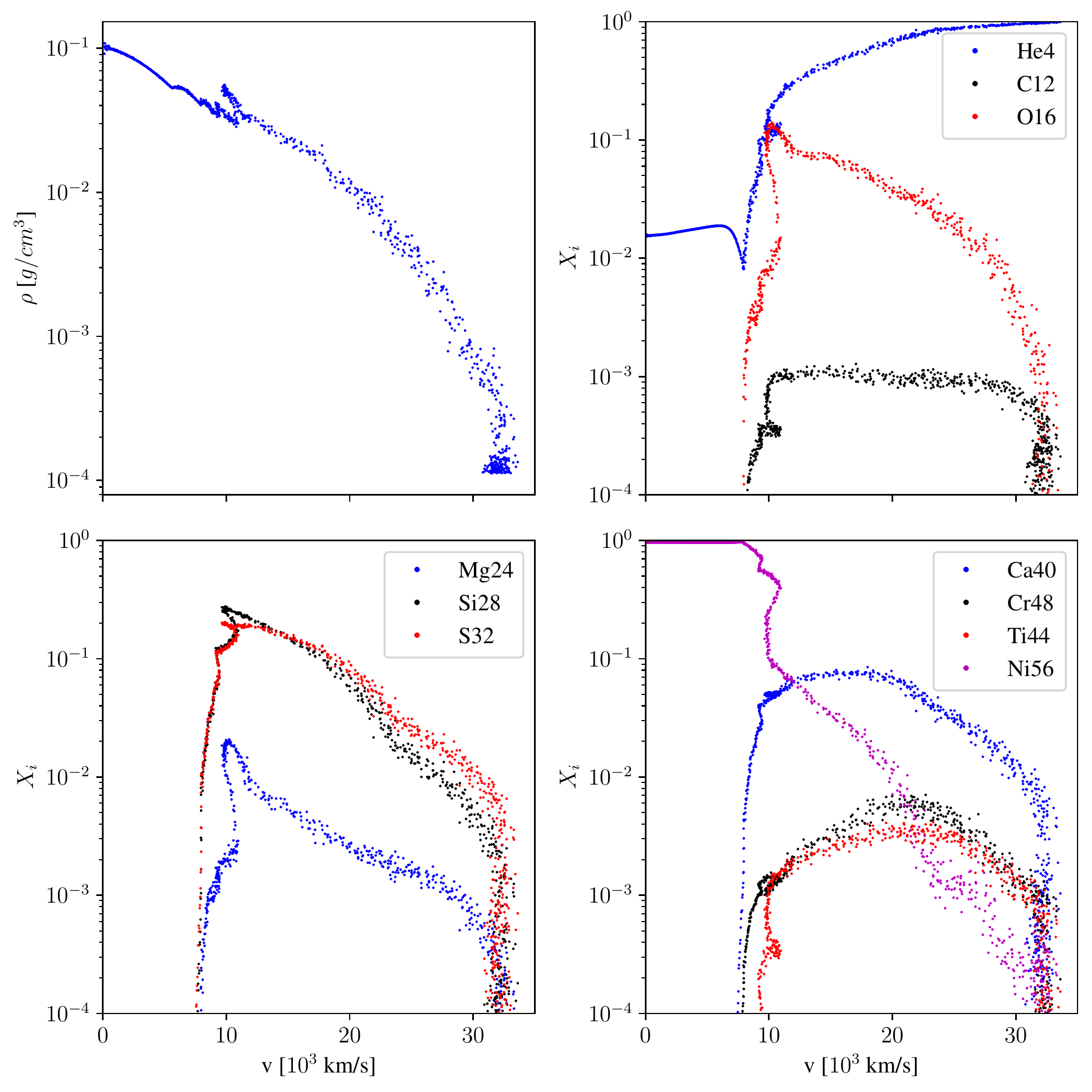}
    \caption{Distribution of densities and compositions in velocity space. This includes a sampling of the left-half plane of Figure~\ref{fig:ejecta}, to avoid the problems of asymmetry around the surviving secondary. Again, we see that $^{56}\mathrm{Ni}$ dominates in the slow-moving Ni bubble in the centre whereas the high-velocity region is dominated by unburnt He4. The higher velocities are also obviously dominated by lower density particles.}
    \label{fig:velocity_space}
\end{figure*}

\begin{figure}
    \centering
    \includegraphics[width=\linewidth, \draft]{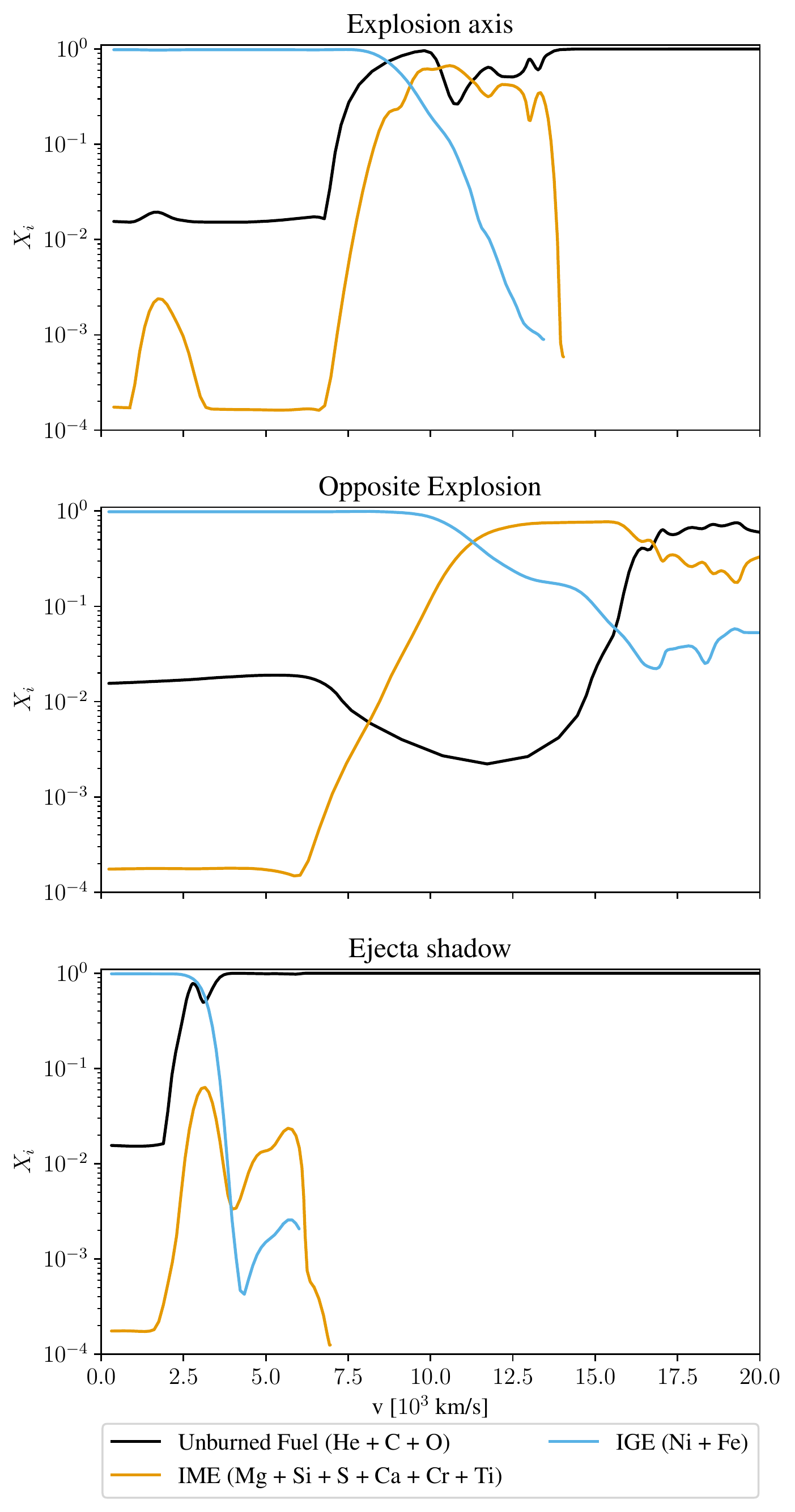}
    \caption{Chemical abundance of intermediate mass elements, iron group elements and unburnt fuel along several axes. These estimates were taken using a cylinder of points oriented in several directions: ``Explosion axis'' points from the centre of the primary prior to explosion to the point of carbon detonation. Despite the significant asymmetry along different axes, all three are characterised by the iron group at low velocities. The intermediate and unburnt fuel is found at intermediate and high velocities, as expected by the lower atomic numbers of these elements.}
    \label{fig:maw_axis}
\end{figure}

We now turn our attention to the characterisation of the ejecta of the edge-lit detonation. This post-detonation period is denoted by the near flat composition lines in Figure~\ref{fig:composition} at $t\gtrsim 222$\,s. The top elemental abundances at late times in the unbound ejecta are: $^{56}\mathrm{Ni}$, 58.6\%; $^{4}\mathrm{He}$, 15.1\%; $^{28}\mathrm{Si}$, 8.92\%; $^{32}\mathrm{S}$, 7.52\%; $^{16}\mathrm{O}$, 2.71\%; $^{40}\mathrm{Ca}$, 2.53\%; $^{36}\mathrm{Ar}$, 2.03\%; $^{52}\mathrm{Fe}$, 0.92\% and $^{24}\mathrm{Mg}$, 0.40\%. Figure~\ref{fig:ejecta} gives a sense of the asymmetry of the ejecta caused by the edge-lit detonation - the central area is dominated by $^{56}\mathrm{Ni}$ produced by burning in the degenerate core while the outer regions contain much more unburned helium. We find the centre of mass (COM) speed of the Ni-56 bubble to be $226$\,\kmpersec, or in vector notation: $[60, 127, 178]$\,\kmpersec. This is low compared to \citet{pakmorFateSecondaryWhite2022}. Our case is most similar to the ``One Explosion'' model of \citet{pakmorFateSecondaryWhite2022}, where the merging of a much higher-mass $0.7$\,\msol \, COHe WD secondary with a $1.05$\,\msol \, primary leads to a detonation of the primary. This results in a $^{56}\mathrm{Ni}$ COM speed of $1\,610$\,\kmpersec. The much higher speed of the COM is likely caused by the much more massive secondary white dwarf and therefore significantly higher orbital velocity of the primary white dwarf adopted in their calculations.

We also consider the yield of the explosion after a period of radioactive decay. Figure~\ref{fig:solar_comparison} presents the elemental abundances relative to iron compared to the same measure for solar composition \citep{asplundChemicalCompositionSun2009}. We assume all isotopes with a half-life shorter than 2\,Gyr have completely decayed. For comparison we include integrated abundances from \citet{gronowDoubleDetonationsSubMCh2021}. \citet{gronowDoubleDetonationsSubMCh2021} experimented with detonations of sub-Chandrasekhar mergers in the mass range between 0.8\,\msol \, and 1.1\,\msol, where the 1.1\,\msol \, COHe primary is considered the closest analogue. These abundances are available on the \textsc{HESMA} database\footnote{\url{https://hesma.h-its.org/}}. The two datasets show many of the same patterns, including a characteristic pattern in and around the transition metals (e.g. $\mathrm{Si}$, $\mathrm{S}$, $\mathrm{Ar}$ and $\mathrm{Ca}$). We can see near-solar production of elements such as $\mathrm{Cr}$ and approximately half-solar production for $\mathrm{Ca}$ and $\mathrm{S}$. Note that the comparison to solar abundances is limited by the fact that this measure is relative to the $\mathrm{Fe}$-content. Since the vast majority of the carbon and oxygen originating in the primary is converted to $^{56}\mathrm{Ni}$, which decays into $^{56}\mathrm{Fe}$, the mass contribution of other elements appears small by comparison. We expect the difference in nickel abundance after decay can be attributed to the presence of $^{22}\mathrm{Ne}$ in Gronow's initial composition; this results in an excess of neutrons (compared to $^{20}\mathrm{Ne}$) which elevate the abundance of neutron-rich iron group isotopes, particularly stable nickel.

\subsection{Network size comparison} \label{sec:network_size}

Although the 55-species network was used to evaluate most quantities, \textsc{AREPO} has the capability to include a larger 384-species network in post-processing. This would be intractable using \textsc{AREPO}'s normal mode, but can be achieved by advecting Lagrangian tracer particles with the flow that record their temperature, pressure and density trajectories. These values can then be used to infer the corresponding 384-species composition using ``Yet Another Nuclear Network'' \citep{pakmorStellarGADGETSmooth2012}. We apply post-processing on one million tracer particles (with this larger 384-species network) from a period just prior to detonation to 100\,s post-detonation. We also apply the 2\,Gyr delay time discussed in the previous section to these tracer compositions for comparison. This is shown in Figure~\ref{fig:network-comparison}.

This comparison makes clear what we might intuitively assume -- the 384-species network is more comprehensive but the 55-species network reproduces it correctly in most ways. Specifically, the $\alpha$-chain is well-reproduced, which accounts for much of the simulated mass. In particular, the match of $^{56}\mathrm{Ni}$ and $^{56}\mathrm{Fe}$ is excellent, so we can be confident in our estimates of the kinematics and energy release from detonation. In terms of isotopic ratios relevant for studies of supernovae, the 55-species network generally performs well but has some weaknesses. The decay chain from $^{54}\mathrm{Zn}$ to  $^{52}\mathrm{Cr}$, which produces $^{52}\mathrm{Fe}$, is not present in the 55-species network -- the 55-species network is thus not capable of simulating the influence of the early UV flux associated with this isotope. The smaller network also fares badly on chemical evolution of the iron group (e.g. $\mathrm{Ni}$, $\mathrm{Mn}$, $\mathrm{V}$). Manganese abundance can be used to differentiate between low and high-density models so the smaller network is also insensitive to these differences. Helium, neon, sulfur and nickel are reproduced to within 20 per cent before decay (similarly; helium, neon, sulfur and iron after decay).

\begin{table*}
\centering
\caption{Yield of the two detonation phases, measured as the difference in mass fraction according to the 55-species network. Note that these changes are based on time and not location. For example helium burning following detonation continues as carbon is ignited. Mass fractions are reported relative to the total amount of mass in the simulated box. Ejecta mass = 1.23\,\msol. Total explosion nuclear energy = $1.61 \times 10^{51}$\,\erg.}
    \label{tab:yield}
    \resizebox{\linewidth}{!}{%
    \begin{filecontents*}{data/table2.txt}
Spec1,Start1,DetI1,DetII1,Final1,Ejecta1,Spec2,Start2,DetI2,DetII2,Final2,Ejecta2
N,5.10e-15,5.45e-15,-1.06e-14,4.78e-18,5.46e-18,Si30,1.22e-08,2.91e-08,3.72e-06,3.76e-06,4.30e-06
P,4.57e-08,1.01e-07,2.48e-04,2.48e-04,2.84e-04,P29,1.34e-07,2.20e-06,2.60e-04,2.62e-04,3.00e-04
He4,2.73e-01,-1.55e-03,-1.47e-02,2.57e-01,1.51e-01,P30,1.28e-07,4.15e-07,6.39e-06,6.93e-06,7.92e-06
B11,5.11e-19,9.85e-19,-1.50e-18,7.44e-22,8.51e-22,P31,4.42e-08,1.44e-07,2.77e-05,2.79e-05,3.19e-05
C12,3.30e-01,-3.16e-03,-3.27e-01,4.86e-04,5.55e-04,S31,1.39e-06,7.72e-06,3.89e-05,4.80e-05,5.49e-05
C13,2.07e-12,6.12e-12,3.92e-11,1.06e-10,6.76e-11,S32,1.38e-04,7.04e-04,6.50e-02,6.58e-02,7.52e-02
N13,7.25e-10,3.92e-09,3.63e-07,3.68e-07,4.20e-07,S33,4.37e-07,3.51e-06,2.50e-05,2.89e-05,3.31e-05
N14,4.96e-10,1.72e-09,-1.63e-09,5.81e-10,6.64e-10,Cl33,2.63e-08,3.77e-07,2.05e-05,2.09e-05,2.39e-05
N15,1.48e-14,9.89e-14,4.78e-09,4.78e-09,5.46e-09,Cl34,6.18e-07,2.55e-06,2.01e-05,2.33e-05,2.66e-05
O15,5.12e-09,5.42e-08,5.66e-07,6.26e-07,7.15e-07,Cl35,2.04e-08,9.27e-08,6.71e-06,6.83e-06,7.81e-06
O16,3.96e-01,1.31e-03,-3.74e-01,2.37e-02,2.71e-02,Ar36,6.20e-05,2.64e-04,1.74e-02,1.78e-02,2.03e-02
O17,1.49e-11,4.75e-11,-6.12e-11,1.14e-12,1.31e-12,Ar37,2.69e-06,1.23e-05,6.80e-05,8.29e-05,9.48e-05
F18,4.88e-12,3.26e-11,-3.10e-11,6.45e-12,7.38e-12,Ar38,1.65e-10,8.81e-10,9.74e-08,9.84e-08,1.13e-07
Ne19,1.59e-11,1.55e-10,1.58e-08,1.59e-08,1.82e-08,Ar39,1.45e-13,3.27e-13,1.93e-12,2.40e-12,2.75e-12
Ne20,5.97e-05,1.24e-04,8.85e-04,1.07e-03,1.22e-03,K39,8.18e-08,2.93e-07,1.07e-05,1.10e-05,1.26e-05
Ne21,1.43e-10,3.09e-10,-3.93e-10,5.89e-11,6.73e-11,Ca40,9.93e-05,4.58e-04,2.16e-02,2.22e-02,2.53e-02
Ne22,5.80e-12,1.32e-11,-1.82e-11,7.32e-13,8.36e-13,Sc43,5.77e-10,5.31e-10,4.55e-09,5.66e-09,6.47e-09
Na22,2.63e-10,6.50e-10,1.99e-08,2.08e-08,2.38e-08,Ti44,9.27e-06,5.77e-05,4.61e-04,5.28e-04,6.03e-04
Na23,1.48e-08,5.42e-08,3.74e-08,1.06e-07,1.22e-07,V47,2.13e-08,2.98e-08,-1.54e-08,3.56e-08,4.07e-08
Mg23,1.09e-07,4.54e-07,4.78e-07,1.04e-06,1.19e-06,Cr48,2.59e-06,2.08e-05,9.69e-04,9.93e-04,1.13e-03
Mg24,8.67e-05,4.82e-04,2.90e-03,3.47e-03,3.97e-03,Mn51,3.90e-09,1.39e-08,6.27e-08,8.05e-08,9.20e-08
Mg25,7.17e-09,1.74e-08,1.84e-07,2.10e-07,2.39e-07,Fe52,1.71e-07,1.59e-06,8.07e-03,8.08e-03,9.23e-03
Mg26,3.94e-09,9.30e-09,1.08e-08,2.40e-08,2.74e-08,Fe56,6.58e-13,9.85e-12,6.68e-06,6.68e-06,7.63e-06
Al25,4.77e-09,3.97e-08,2.49e-06,2.53e-06,2.89e-06,Co55,1.51e-10,1.13e-09,1.36e-07,1.38e-07,1.57e-07
Al26,2.83e-07,8.20e-07,2.31e-06,3.42e-06,3.90e-06,Ni56,1.86e-09,2.00e-08,5.13e-01,5.13e-01,5.86e-01
Al27,1.69e-07,2.33e-07,3.93e-06,4.33e-06,4.95e-06,Ni58,4.39e-13,3.49e-12,6.71e-03,6.71e-03,7.66e-03
Si28,1.79e-04,1.27e-03,7.65e-02,7.80e-02,8.92e-02,Ni59,1.43e-12,1.35e-11,9.66e-04,9.66e-04,1.10e-03
Si29,9.62e-08,2.46e-07,7.62e-05,7.66e-05,8.75e-05,-,-,-,-,-,-
\end{filecontents*}
\pgfplotstabletypeset[
    multicolumn names,
    col sep=comma, 
    columns={Spec1,Start1,DetI1,DetII1,Final1,Ejecta1,Spec2,Start2,DetI2,DetII2,Final2,Ejecta2},
    columns/Spec1/.style={string type, column name=Species, column type={l|}}, 
    columns/Start1/.style={column name=Start, 
    column type={l},
    string type}, 
    columns/DetI1/.style={column name=He det, 
    column type={l},
    string type}, 
    columns/DetII1/.style={column name=C det, 
    column type={l},
    string type}, 
    columns/Final1/.style={column name=Final comp.,
    column type={l|},
    string type}, 
    columns/Ejecta1/.style={column name=Ejecta composition,
    column type={|l},
    string type}, 
    columns/Spec2/.style={string type, column name=Species, column type={l|}}, 
    columns/Start2/.style={column name=Start, 
    column type={l},
    string type}, 
    columns/DetI2/.style={column name=He det, 
    column type={l},
    string type}, 
    columns/DetII2/.style={column name=C det, 
    column type={l},
    string type}, 
    columns/Final2/.style={column name=Final comp.,
    column type={l|},
    string type}, 
    columns/Ejecta2/.style={column name=Ejecta composition,
    column type={|l},
    string type}, 
    every head row/.style={before row=\toprule, 
    after row=\midrule}, 
    every last row/.style={after row=\bottomrule},
    ]{data/table2.txt}
    }
    
\end{table*}

\begin{table}
\centering
\caption{Mass composition of the surviving secondary $\approx 400$\,\s \, after detonation for the T-120 model. Note that the mass is dominated by $^{4}\mathrm{He}$, with trace amounts of $^{56}\mathrm{Ni}$. (This composition was computed using \textsc{AREPO}'s internal composition rather than the post-processing discussed in Section~\ref{sec:network_size}.)}
    \label{tab:secondarytab}
    \resizebox{0.7\linewidth}{!}{
    \begin{filecontents*}{data/table3.txt}
Spec1,Xnuc1,Spec2,Xnuc2
N,3.43e-21,Si30,2.98e-09
P,1.02e-06,P29,2.95e-07
He4,2.19e-01,P30,4.26e-08
B11,3.49e-26,P31,1.20e-08
C12,2.72e-05,S31,2.01e-08
C13,1.02e-10,S32,7.90e-06
N13,5.98e-08,S33,9.50e-08
N14,5.17e-12,Cl33,1.59e-07
N15,3.03e-13,Cl34,2.82e-08
O15,3.60e-11,Cl35,1.46e-09
O16,3.59e-07,Ar36,2.11e-06
O17,1.79e-16,Ar37,2.80e-09
F18,4.65e-15,Ar38,2.52e-11
Ne19,2.05e-12,Ar39,1.20e-15
Ne20,5.16e-07,K39,1.20e-09
Ne21,2.96e-15,Ca40,2.69e-06
Ne22,2.71e-16,Sc43,9.42e-13
Na22,5.07e-12,Ti44,7.34e-08
Na23,2.62e-12,V47,1.57e-12
Mg23,4.12e-12,Cr48,1.86e-07
Mg24,4.87e-07,Mn51,1.08e-12
Mg25,8.34e-09,Fe52,1.97e-06
Mg26,1.12e-11,Fe56,1.46e-08
Al25,1.02e-07,Co55,1.77e-12
Al26,7.65e-09,Ni56,1.73e-03
Al27,2.39e-09,Ni58,3.13e-05
Si28,9.26e-06,Ni59,4.48e-06
Si29,5.29e-08,-,-
\end{filecontents*}
\pgfplotstabletypeset[
    multicolumn names,
    col sep=comma, 
    columns={Spec1,Xnuc1,Spec2,Xnuc2},
    columns/Spec1/.style={string type, column name=Species}, 
    columns/Xnuc1/.style={string type, column name=M [\msol], column type={l|}}, 
    columns/Spec2/.style={string type, column name=Species}, 
    columns/Xnuc2/.style={string type, column name=M [\msol]}, 
    every head row/.style={before row=\toprule, after row=\midrule}, 
    every last row/.style={after row=\bottomrule},
    ]{data/table3.txt}
    }
\end{table}

There is also a characteristic ``ejecta shadow'' formed by the secondary. The unbound ejecta have a total mass of 1.23\,\msol \, and move with a COM speed of $140$\,\kmpersec, $[ -42, -133, -1]$\,\kmpersec. Figure~\ref{fig:maw_axis} attempts to characterise the asymmetry of the ejecta cloud by showing the differences in chemical abundance along several axes. Since the ``Explosion axis'' passes through the ejecta shadow, the $^{56}\mathrm{Ni}$ bubble is interrupted by a dip to low atomic weights in this direction. While the approximate shape of the graph is similar in all three cases, the gradient of the drop-off is different for all three cases. This may be an observable that could be measured on Earth, particularly if the line of sight passes through the ejecta shadow. 

Figure~\ref{fig:velocity_space} shows the distribution of densities and compositions in velocity space (considering only particles in the left half of the simulated space to avoid the asymmetry due to the secondary). As we would expect, higher velocities are dominated by lower densities and particularly unburned $^{4}\mathrm{He}$, while the heavy $^{56}\mathrm{Ni}$ bubble dominates the low-velocity space. We can estimate the total kinetic energy released by identifying all the unbound cells in the simulation long after the detonation and finding the sum of their kinetic energies. This gives $1.19 \times 10^{51}$\,\erg, which is consistent with a SN\,Ia explosion. Similarly, we can estimate the release of nuclear energy by comparing the proportions of all species at the beginning and end of the simulation to estimate the mass deficits for each species -- this gives $1.61 \times 10^{51}$\,\erg. Again, our results are consistent with those of  \citet{pakmorFateSecondaryWhite2022} for their ``One Explosion'' model -- $1.4 \times 10^{51}$\,\erg\ and an ejecta mass of $1.09$\,\msol. 

\subsection{Fate of the Secondary} \label{sec:secondary_fate}

We now consider the fate of the secondary 100\,s after detonation time. At this time, we find that the secondary has survived and retains $0.22$\,\msol \, of bound mass, moving at $1\,830$\,\kmpersec ($[ 960, 1560, 9]$ \kmpersec). We can compare this with the linear orbital velocity just before detonation ($[1170,  1371, 0]$ \kmpersec, $v = 1800$\,\kmpersec) to compute the kick velocity. This gives $[273, -124, -12]$\,\kmpersec, $v = 301$\,\kmpersec. \citet{shenThreeHypervelocityWhite2018} predict that in ``dynamically driven double-degenerate double-detonation'' the companion WD survives the explosion and is flung away with a velocity equal to its $> 1\,000$\,\kmpersec pre-SN orbital velocity. However, the survival of the secondary is potentially in conflict with the small number of candidates for surviving secondary WDs in the Solar neighbourhood \citep{shenThreeHypervelocityWhite2018}.

There are other important questions related to the survival of the secondary. For example, can we conclude that the surviving secondary is still a WD? Comparisons of the initial state of the secondary show that the central density has decreased from approximately $7.51 \times 10^5$\,\gcmcubed\ to $1.5 \times 10^5$\,\gcmcubed. The temperatures and densities of the remaining bound particles confirm that the secondary lies well within the degenerate regime \citep[e.g., see Figure 8.1 of][]{schwarzschildStructureEvolutionStars2015}, indicating that it remains a WD. However, there is some enrichment of the secondary with heavier elements. The total composition of the bound particles making up the surviving secondary includes approximately 99.2 per cent $^4$He and 0.8 per cent $^{56}\mathrm{Ni}$, which might be relevant for the further evolution of the surviving secondary via heating produced by radioactive decay \citep{shenWaitItPostSupernova2017}. A detailed breakdown of the secondary's mass composition $400$\,\s \, post-detonation is given in Table~\ref{tab:secondarytab}. This is a relatively high level of pollution which could be observed, particularly given these heavier elements are concentrated in the atmosphere. Evolving this structure using a stellar evolution code such as MESA to find these observables at different evolutionary stages following the explosion would be a very interesting exercise to pursue but beyond the scope of this paper.

\citet{tanikawaDoubledetonationModelsType2019} presents a suite of SPH merger simulations that provide a useful comparison in terms of the initial and final secondary composition. Their ``He45'' model consists of a $1.05$\,\msol primary (including the mass of a helium shell) coupled with a $0.45$\,\msol helium secondary. This pair detonates via the D6 mechanism, producing a surviving companion that captures $0.02$\,\msol. Comparing with the non-He elements in Table~\ref{tab:secondarytab}, we obtain a captured mass of $1.77 \times 10^{-3}$\,\msol. While this is a far lower number, this is unsurprising given that a significant portion of our secondary's mass had already been accreted onto the primary. Our figure is closer to the analytic estimate of \citet{shenWaitItPostSupernova2017}:  $0.006$\,\msol of $^{56}\mathrm{Ni}$ bound to a secondary of $0.3$\,\msol.

\subsection{Inspiral time} \label{sec:inspiral}

\begin{figure*}
    \centering
    \includegraphics[width=0.8\linewidth, \draft]{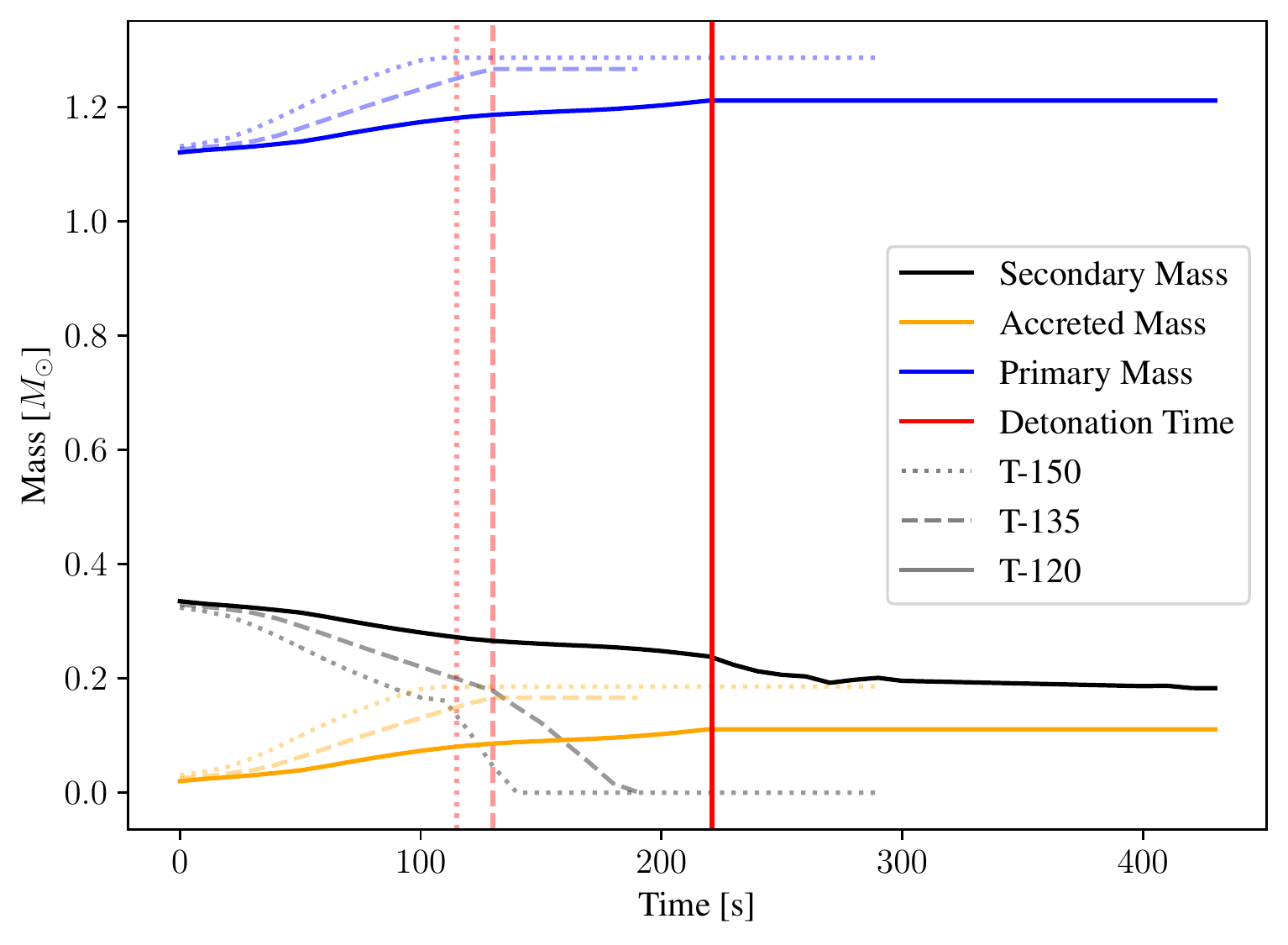}
    \caption{These data are generated using the particle labels to identify bound particles originating from the secondary. Particles which originated in the secondary but are primarily under the gravitational influence of the primary are classified as accreted. The area of gravitational influence is delineated by the radius of the Lagrange point between the two stars.}
    \label{fig:accretion_rate}
\end{figure*}

\begin{figure}
    \centering
    \includegraphics[width=\linewidth]{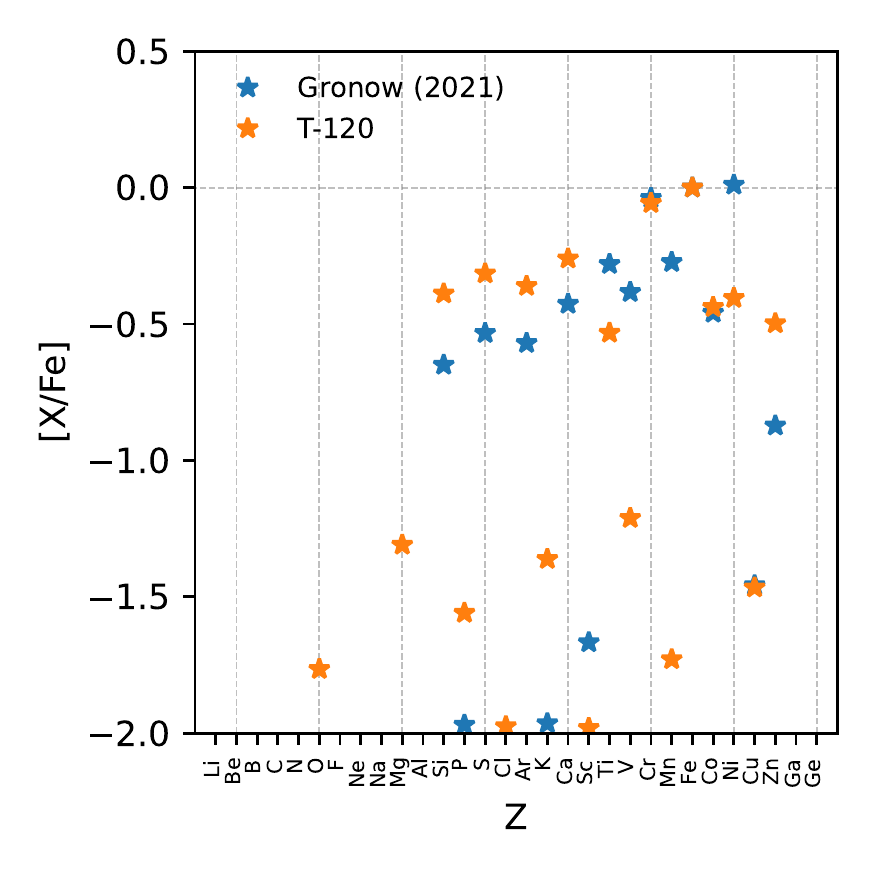}
    \caption{Comparison of elemental composition of the ejecta relative to iron and solar composition 2\,Gyr after the explosion \citep{asplundChemicalCompositionSun2009}, i.e. taking into account radioactive decay of unstable isotopes in the ejecta. ``Gronow (2021)'' refers to the 1.1\,\msol \, COHe WD model in \citet{gronowDoubleDetonationsSubMCh2021}}
    \label{fig:solar_comparison}
\end{figure}

As previously discussed, \textsc{INSPIRAL} is used to accelerate the loss of angular momentum so that the outcome of a merger can be evaluated in a logistically-tractable amount of time. In this section, we justify the use of this technique. One way to do this is to compare the outcomes of the three scenarios listed in Table~\ref{tab:inspiral-comparison}. The amount of inspiral time does indeed have a dramatic effect on the merger, particularly as related to the secondary. For Model T-150, the secondary is completely disrupted slightly before the detonation occurs. For Model T-135, the shockwave dissipates much of the remaining mass, creating an object that is weakly gravitationally bound and strongly asymmetrical. The different amounts of accreted matter on the primary will translate to different explosion energies. A comparison of the accretion profiles of the three cases is shown in Figure~\ref{fig:accretion_rate}, with T-120 shown in solid colours. This plot was created using the Passive Scalar labelling discussed in Section~\ref{sec:setup}; we do this by finding all bound cells labelled as originating in the secondary and finding the collinear Lagrange point between the centres of the two WDs. Bound cells closer to the primary centre have been accreted whereas those closer to the secondary centre still belong to the secondary. Note how the mass accretion rates (i.e. gradients) differ significantly between the three cases.

Although the accretion patterns are significantly different, the detonation characteristics in the primary are not. While all three detonations occur at different times, they begin at the base of the Helium layer in regions with similar values for the primitive variables compared to those discussed in Section~\ref{sec:detonation}. Similarly, once the helium detonation has begun, a carbon detonation follows within $0.3$ -- $0.4$\,\s \, for each case. We can conceive of the chosen inspiral time as a parameter of the system which selects for a particular amount of angular momentum to be contained in the system, meaning that each of the cases is a realistic representation of the outcome for that particular angular momentum. However, we consider the cases with the longest practical merger times (T-113 to T-120) to be the most realistic because in these models the time of the accelerated phase of angular momentum loss is the minimum required to keep the simulations tractable. In reality, the system will ideally go through the same stages as it does in the simulation by on much longer timescales. We shall use the T-120 model for illustrative and comparison purposes. We discuss the realism of this system further in the following section. Longer acceleration times would be closer to a violent merger situation while shorter times do not result in interesting dynamics in a computationally feasible amount of time. Accumulated numerical errors also caused issues in such cases. For example, simulations with shorter acceleration times than model T-110 resulted in leakage of helium into the WD interior. 

\subsection{Progenitor system} \label{sec:progenitor}

The mass and chemical structure of the two WDs before merging are the outcome of their earlier formation history which includes several phases of binary interactions \citep{ruiterBrightnessDistributionType2013}. A DD system consisting of a 1.1\msol\, CO WD and a 0.35\msol\, HE WD can be achieved from the evolution of two stars of intermediate-mass on the main sequence. We have investigated the possible evolutionary paths prior to their merging by following the evolution of $10^7$ binaries up to an age of 13\,Gyrs using the rapid binary star evolution code \textsc{BSE} \citep{hurleyEvolutionBinaryStars2002, hurleyFormationBinaryMillisecond2010}. The primary and secondary masses were taken in the range $0.8-10$\,\msol\, and initial orbital periods in the range $10-10\,000$\, days. The masses of the primaries were randomly chosen according to Kroupa's initial mass function \citep{kroupaVariationInitialMass2001} and those of the secondary stars according to a flat mass distribution with $q=M_2/M_1\leq 1$ \citep{ferrarioConstraintsPairingProperties2012}. The initial period distribution was assumed to be uniform in the logarithm \citep{kouwenhovenPrimordialBinaryPopulation2007}. In our calculations we have used complete common envelope efficiency ($\alpha_{\rm CE}=1$) to be consistent with previous studies of the predicted rate of SNe\,Ia from the DD merger channel \citep{ruiterDelayTimesRates2011,ruiterFormationNeutronStars2019}.

We then extracted all single WDs that resulted from the merging of two WDs. The most common evolutionary path that leads to these events usually requires two common envelope phases. If the stars do not merge during common envelope evolution, they form a close binary system consisting of two WDs. 

\begin{figure*}
    \centering
    \includegraphics[width=0.85\linewidth ]{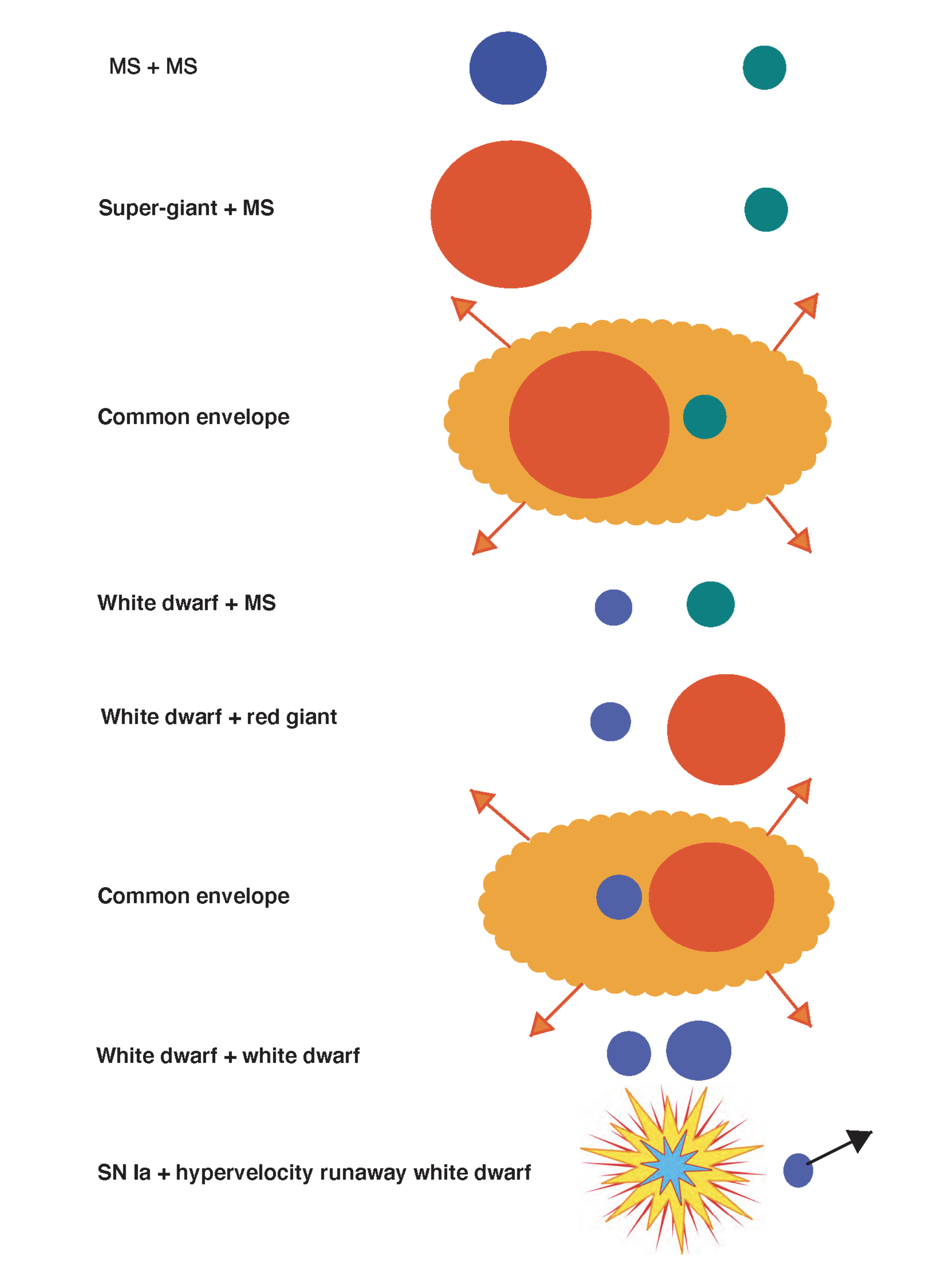}
    \caption{Schematic diagram of the evolutionary path to SN\,Ia explosion with hypervelocity runaway WD. }
    \label{fig:BinaryEvolution}
\end{figure*}

\begin{figure*}
    \centering
    \includegraphics[width=\linewidth]{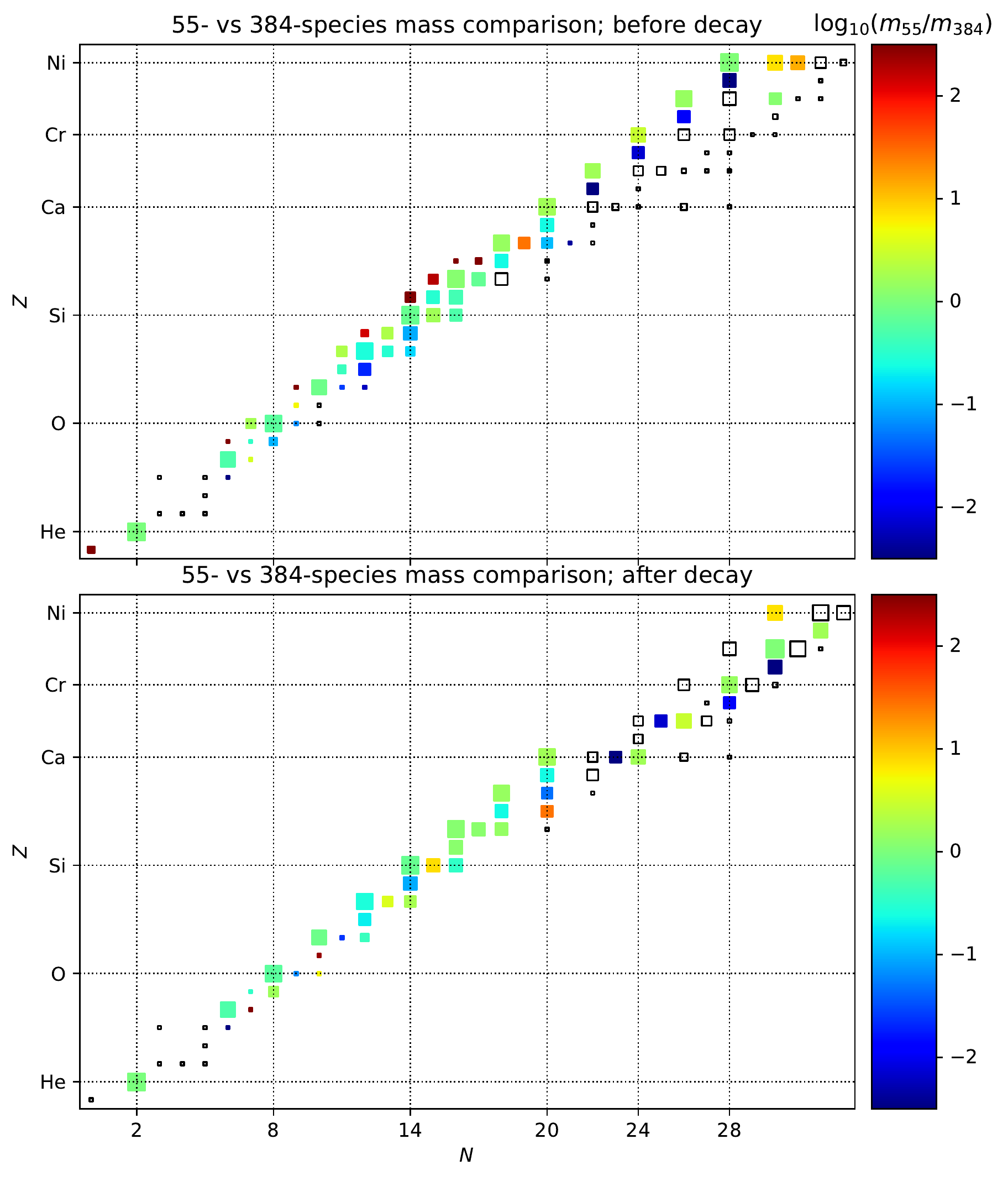}
    \caption{Mass comparison of the \textsc{AREPO} simulation using the 55-species network, compared to the post-processed result using 384-species. The top panel shows the relative abundance of the various elements approximately 100\,s after detonation, whereas the bottom panel compares the two after 2\,Gyr. Clear boxes represent elements which are present in the 384-species network but absent in the 55-species network. Larger box sizes indicate a larger contribution to the total mass in the simulation.}
    \label{fig:network-comparison}
\end{figure*}

We have found that the most common initial setup on the zero age main sequence consists of a primary star with a mass in the range $5-7$\,\msol\, and a secondary with a mass in the range $2-3$\,\msol\, orbiting each other on a period of $500-700$\,days. The key evolutionary points can be sketched as follows. First the primary evolves to become a supergiant, with a massive CO core, on the asymptotic giant branch. As the supergiant fills its Roche lobe unstable mass transfer begins and a common envelope forms. This phase drastically reduces the separation between the two stars and strips the primary of its envelope turning it into a naked helium star that evolves into a massive CO WD with an outer helium shell and no hydrogen left. In the meantime, the secondary runs out of hydrogen in its core, expands to become a red giant, and fills its Roche lobe. At this point another common envelope phase, this time initiated by the secondary, begins. Because helium has not been ignited in the core and the envelope of the secondary is ejected during common envelope, the star emerges from this evolutionary phase as a very low-mass helium star whose interior will never reach the necessary density and temperature conditions to burn helium and thus evolves into a HE WD. The typical final product consists of a massive CO WD primary ($1-1.2$\msol) and a low-mass HE WD secondary ($0.2-0.35$\msol) orbiting each other at a separation of some $\times 10^9$\,cm ($\sim 1.44 \times 10^{-2} \, R_{\odot}$). These parameters are in close agreement with those employed for our super-Chandrasekhar merger at the start of our simulations and correspond to the point when we switch on the accelerated in-spiralling phase (see \ref{sec:inspiral}) to make the merger simulations tractable. A diagrammatic sketch of the evolutionary path is shown in Figure \ref{fig:BinaryEvolution}.

While conventional arguments may suggest that a binary with WD masses of $1.1$\msol \, and $0.35$\msol \, ($q=0.32$) could encounter stable mass transfer and avoid a merger \citep[e.g.,][]{marshMassTransferDouble2004, danHowMergerTwo2012}, a merger outcome is not unlikely. After accretion starts, the donor star would likely then further overfill its Roche lobe (as a result of mass loss through nova mass ejection), driving the binary toward a smaller separation resulting in a merger \citep{shenEveryInteractingDouble2015}. While further, more detailed studies are required to assess the specific outcome of closely interacting WDs, if systems such as those studied here are indeed prone to merging, this could easily explain why a large number of stably-accreting systems hosting lower mass, helium-rich WDs are not observed \cite[e.g., ][]{brownMostDoubleDegenerate2016}. 

\subsection{Comparison to isothermal, constant composition model} \label{sec:isothermal_comparison}

In order to assess the sensitivity of the merger outcome on the composition profile, we compare the result of our WDEC merger to two constant-composition models similar to those found in the literature. The ``Bare Core'' model was generated using the hydrostatic equilibrium integration method with the same target mass and central density as the WDEC model (i.e. total mass 1.1\,\msol, and 70 per cent O16, 30 per cent C12 was used in each of its cells) and a constant temperature of $T = 5 \times 10^5$\,K. The ``Thick He Layer'' model is identical, except its outer radii (beyond $4 \times 10^8$ \cm) have been set to be 100 per cent helium, corresponding to 1.4 per cent of the total mass. Note that the pressure and density values of these cells remained the same as the \textit{Bare Core} case, so this introduces Helium at much higher pressures and densities than those that would be immediately accreted on the bare core. The purpose of this comparison was to see whether the inclusion of Helium at high pressure is vital to initiating the detonation. Both these models were transformed into a 3D structure using HEALPIX and evolved first in isolation to check for stability. The normal pure $^{4}\mathrm{He}$ background was used with density $\rho_v = 1 \times 10^{-4}$\,\gcmcubed. The merger was simulated using the inspiral parameters in Table~\ref{tab:inspiral-comparison}, T-120, to match the conditions placed on the WDEC model as closely as possible. The same $0.35$\,\msol \, secondary was used for all three cases as well.

\begin{landscape}
\begin{figure}
    \centering
    \includegraphics[width=\linewidth,\draft]{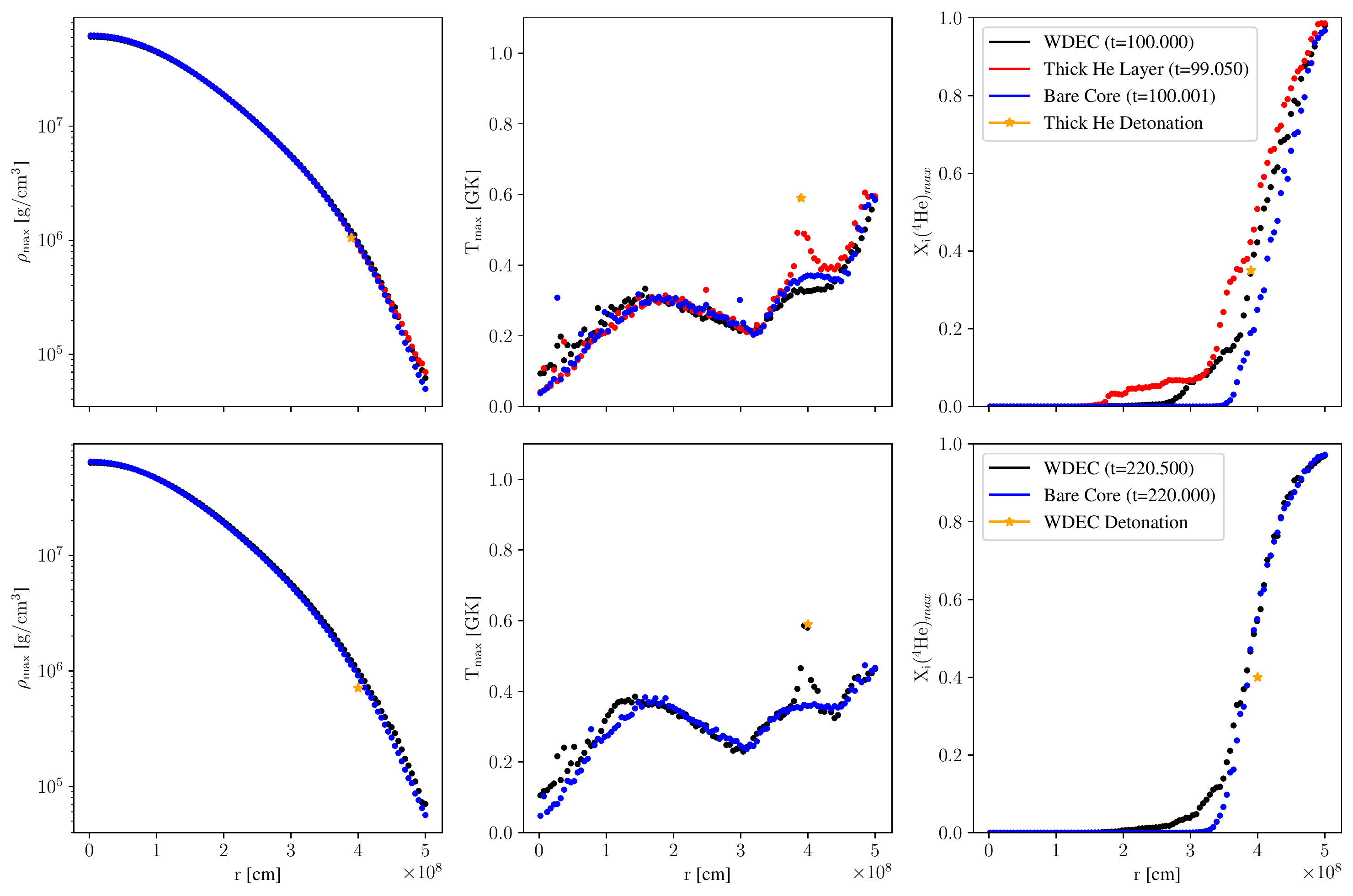}
    \caption{Comparison of peak helium abundance, density and temperature as a function of radius for the three WD structures (see text).Top row:  This effectively shows the penetration of Helium to the interior of the WD in each of the three models. Bottom row: the orange star indicates the conditions for just prior to helium detonation discussed in Section~\ref{sec:detonation}. We observe that models which detonate have a sizable helium content and high temperature at a radius of approximately $4 \times 10^8$ \cm \, compared to those that don't, and that the model with the larger helium layer reaches this point sooner.}
    \label{fig:model_comp}
\end{figure}
\end{landscape}

The \textit{Bare Core} structure was left to evolve until the secondary was almost completely accreted and no double detonation was observed (up to $t = 400$\,\s). Similarly to the \textsc{WDEC} model, this model produces some heavier elements such as $^{20}\mathrm{Ne}$, $^{24}\mathrm{Mg}$ and $^{28}\mathrm{Si}$, but burning in the accreted helium region could not trigger a carbon detonation. By contrast, the CO WD with a \textit{Thick He Layer} detonated after only $100$\,\s \, in a manner similar to the previous models that used the \textsc{WDEC} structure for the primary. Figure~\ref{fig:model_comp} illustrates the difference in the \textit{Bare Core} model in comparison to the other two; this figure shows the peak density, temperature and pressure of cells in each primary interior as a function of their Helium composition. While all three models had very similar temperatures across this range, note that the \textit{Bare Core} structure lies strictly below the other two curves in terms of density and pressure. This suggests only the \textsc{WDEC} and \textit{Thick He Layer} models contained cells rich in Helium at high enough temperatures and pressures to trigger ignition. The \textsc{WDEC} model relies on accretion of mass to raise the density and pressure at the bottom of the Helium layer while the \textit{Thick He Layer} model has these values already sufficiently high to reach triggering conditions after a small amount of helium is accreted. This vast difference in outcomes reinforces the need to use accurate chemical profiles for merger simulations because EOS estimates of resultant temperature, pressure and density are highly sensitive to precise proportions of species, so there is significant advantage to be gained by adding more accuracy to this element of the model. One limitation of the chemical structures produced by \textsc{WDEC} is that only evolution of isolated WDs is considered. 

\section{Conclusions} \label{sec:conclusions}

This paper discusses some of the important research questions associated with the simulation of physically realistic WD mergers. Better understanding of mergers will have downstream consequences for our understanding of, for instance, supernova type Ia, the formation of highly magnetic compact stars, and the chemical enrichment of galaxies. 

We have used the WD structure code \textsc{WDEC} to create a variable composition and temperature 1D WD structure which was ported into the time evolution code \textsc{AREPO} using a HEALPIX mapping. Our simulations have showed detonations which are consistent with the edge-lit scenario, creating an asymmetric ejecta pattern compatible with an off-centre explosion. The ejecta that are flung into space are dominated by the presence of $^{56}\mathrm{Ni}$, $^{4}\mathrm{He}$, $^{28}\mathrm{Si}$, and $^{32}\mathrm{S}$.

The \textsc{INSPIRAL} routine was used to accelerate the timescale of the merger by increasing the rate of loss of angular momentum which did not significantly affect the detonation mechanism. Although the amount of mass accreted, explosion energy, and fate of the secondary are dependent on the \textsc{INSPIRAL} parameters, we have argued that the slowest accretion Cases (T-113 to T-120) are the most realistic, since the sampling of the phase space of states corresponds more closely to that experienced by the primary and secondary over a much longer co-rotating period. Instead, the violent merger cases (T-135 and T-150) caused large amounts of mass to be deposited very quickly onto the primary. We caution that aggressive inspiral times can potentially lead to unphysical results. Furthermore, the huge tidal stresses either ejected larger amounts of mass in the tidal tail or a large amount of mass was blown away during detonation. We find that in T-113 to T-120, the merger survives as a $0.22-0.25$\,\msol \, degenerate remnant moving at $>1\,700$ \kmpersec, consistent with the characteristics of the candidate stars found by \citet{shenThreeHypervelocityWhite2018} and comparable to the velocities given in \citet{pakmorFateSecondaryWhite2022}. Comparisons of the initial state of the secondary show that although its central density has decreased and that it is now a much bloated star, it remains degenerate. It is enriched by metals with about 0.8 per cent $^{56}\mathrm{Ni}$ by mass, concentrated at the surface. It is unclear what its evolutionary path, and stable structure, might be. This aspect may warrant further investigation in a future paper.

The results from two isothermal, constant composition models were compared to the results of our \textsc{WDEC} T-120 detonation. These all used the same mass, central chemical composition, and \textsc{INSPIRAL} parameters. One isothermal structure, the \textit{Bare Core} model, had no helium layer whilst the other, the \textit{Thick He Layer} model, had a pure helium layer for $r > 4 \times 10^8$ \cm. The \textit{Bare Core} was left to evolve until the secondary had been almost completely accreted and no detonation was observed. Instead, the \textit{Thick He Layer} model exploded after only $t = 100$\,s.  We feel this reinforces the role of accurate chemical, density, pressure, and energy profiles for the much more massive, accreting WD in double-degenerate merger simulation. 

Upcoming simulations using \textsc{AREPO} will allow us to study other aspects of the time dynamics of mergers, such as different masses and mass ratios \citep[e.g. see][]{danStructureFateWhite2014}. In a follow-up work we plan to investigate the conditions for detonations, amount of matter ejected, and the fate of the secondary star for a wide range of WD masses and mass ratios similarly to the studies of \citet{danHowMergerTwo2012, danStructureFateWhite2014}, but at much higher resolutions. Neglecting the presence or the possible formation of magnetic fields during merging event may also be a limitation of this kind of work. This is because strong magnetic fields are likely to be generated due to the $\alpha-\Omega$ mechanism that could be triggered by the onset of strong differential rotation and seed fields \citep[e.g.][for a review on magnetic field generation see \citet{ferrarioMagneticFieldGeneration2015}]{duncanFormationVeryStrongly1992, thompsonNeutronStarDynamos1993, wickramasingheMostMagneticStars2014}. The $\alpha-\Omega$ mechanism has been suggested to be responsible for the origin of strong magnetic fields ($10^6$ to $10^9$\,G) in approximately 10 per cent of isolated WDs \citep{ferrarioMagneticFieldsIsolated2020}. Therefore, a proper treatment that includes MHD effects may shed more light on the physics that governs stellar merging events. 

\section*{Code Availability}
The \textsc{AREPO} code is publicly available \href{https://arepo-code.org}{here}, however some features are only available on the private "Development" branch. Access to the development branch is subject to approval from \textsc{AREPO}'s development team. The WDEC code is publicly available \href{https://github.com/kim554/wdec}{on Github}. The MESA code
is publicly available \href{http://mesa.sourceforge.net}{here}.

\section*{Software}
This work made use of the following open-source software projects: NumPy \citep{harrisArrayProgrammingNumPy2020}, MatPlotLib \citep{hunterMatplotlib2DGraphics2007}, SciPy \citep{virtanenSciPyFundamentalAlgorithms2020}.

\section*{Facilities and Acknowledgements}
This research was undertaken with the assistance of resources and services from the National Computational Infrastructure (NCI), which is supported by the Australian Government. Computation time was contributed by the Australian National University through the Merit Allocation Scheme (ANUMAS), the National Computational Merit Allocation Scheme (NCMAS) and the UNSW HPC Resource Allocation Scheme. A.J.R. has been supported by an Australian Research Council Future Fellowship through award number FT170100243. U.P.B was supported by the Australian Government Research Training Program (AGRTP) Fee Offset Scholarship and ANU PhD Scholarship (Domestic).

\section*{Data Availability}
The data underlying this article will be shared on reasonable request to the corresponding author.


\bibliographystyle{mnras}
\bibliography{main}






\bsp	
\label{lastpage}
\end{document}